\documentclass[fleqn]{article}
\usepackage{amscd,amsmath,amssymb}
\usepackage[dvips]{graphicx}
\title{$\mathfrak{P}$-STRUCTURES, $\mathfrak{P}$-GEOMETRY  AND OBSERVER'S PERCEPTIVE SPACE}
\author{Anna Astakhova${}^\ast$, Kyrill Goods and Sergey
Kokarev${}^\dag$}
\date{${}\ast$Baumann State Technical University, Moscow, Russia\\      %%   different addresses +)
           ${}^\dag$Regional Scientific-Educational Center "Logos", r.22. Respublikanskaya 80,
Yaroslavl, 150000, Russia\\ tel/fax: (4852)726346}

\begin{document}
%\twocolumn[

\maketitle

\begin{abstract}
    {Abstract mathematical theory of an observer is elaborated upon the basis of
A.Poincare's ideas on the nature of geometry and the role of observer's perceptive space.
The said theory is generalizing reference frames  theory in GR.
Physical structure ($\mathfrak{P}$-structure) and corresponding physical geometry
($\mathfrak{P}$-geometry) notions, representing properties invariance of some physical
objects and their relations, are introduced.
$\mathfrak{P}$-structure of classical physical time and its corresponding chronogeometry
is considered as an example. Some quantitative characteristics of observer's visual
space geometry are experimentally determined.
The affine model of visual geometry is offered to interpret experimentally sampled data.
The connection of the obtained results with some problems of theoretical physics
is being discussed.}
\end{abstract}
%]

\section{Introduction}

The creation of special relativity (SR) and general relativity (GR)
has been a powerful incentive for applying geometrical ideas to the basement
of modern theoretical physics \cite{geom1,geom2,geom3,geom4,geom5}.
No physical theory pretending at present to describe nature's fundamental
ground principles is free to certain extent from geometrical ingredients.
Though as our knowledge about physical world structure is deepening aided
by accelerators and mighty telescopes, detail description is getting more and more
intricate and refined, geometrical ideas and methods, taking shape of more modern
and abstract forms,  remain basic working language of theoretical and mathematical physics.

At the beginning of the last century A.Poincare,
possessing ingenious geometrical intuition, was one of the first to question the nature
of geometry's axioms \cite{poincare}.
Analyzing  geometries known at that time, their origin and correlation with experience,
he has come to conclusion that geometry was an artifact of physical objects
properties.
He wrote in particular:

\medskip
\noindent \it "Where do primary geometry principles originate from? Are they prescribed by logic?
Lobachevsky, having created non-Euclidean geometries, proved it wrong.
Don't we discover space with our perception? Again, not, since the space of our perception
can teach us is totally different from that of geometrist's.
Does geometry stem from experience at all?
Profound research shows us that it's not so.
We can conclude from here that these principles are no more than conditional \dots
and if we were trans\-por\-ted into some other world
(I call it non-Euclidean and am trying to depict it),
then we would draw our attention to other foundations"\, \rm \cite[p.10]{poincare}.

\medskip
%Appropriate perception of  Poincare's ideas, that in our opinion have substantially
%overtaken their time and remain  unappreciated up to now, has been  hindered  by philosopher's
%much too active interference. Authorship of direction in philosophy
%called "conventionalism"\, traditionally connected with Poincare,
%can be to greater extent assigned to philosophers-interpreters
%(including Marxist philosophers) of mathematician's creative heritage.
%As to Poincare, he has only written about geometry being conditioned by a class of physical
%objects to be described --- the idea is close in its spirit to GR.
%{\it "If we now turn to the question whether Euclidean geometry is true,
%we will find out that it doesn't make any sense.
%It would be the same as asking which system is true --- metric or system with ancient measures,
%or which coordinates are right  --- Cartesian  or polar.
%No geometry can be righter than any other.
%It's a matter of convenience which one to choose"}
%\cite[p.10]{poincare}.
\noindent Thus, according to Poincare's thought, different
geo\-met\-ri\-es are different languages expressing some invariant
physical sense which, as a matter of fact, is an object of
physicists' and mathematicians' researches.

As Poincare's profound analysis has shown, observer's visual space,
being a subspace of broader perceptive space,
plays exceptional role in geometry axioms for\-ma\-ti\-on and is by its substantial
properties different from geo\-met\-ri\-cal space.

\medskip
\noindent \it "The space of imagination is just an image of geometrical space ---
an image changed by some sort of prospect;
we do not imagine therefore external bodies in geometrical space,
but we are speculating  about these bodies as if they were placed in geometrical
space"\, \rm \cite[p.10]{poincare}.

\medskip
Developing this Poincare's thought, it can be assumed that there exists some observer's
inner geometry  which turns out to be imperceptibly built into our observations,
laws, equations, and physical principles.
In fact, this is not, if understood in broader sense, a new thought
and was already developed in sound perception laws discovered XIX century.
So, famous Weber-Fechner law, stating that perceived (seeming)
sound signal force is proportional to logarithm of its true force,
actually asserts non-Euclidean nature of  hearing subspace of an observer's perceptive space.
Since an observer receives more than 80\% of information  through  eyesight,
ana\-lo\-go\-us researches of an observer's visual space are of interest.
Information on quantitative characteristics of visual
(or more general --- perceptive) space can be used in a wide class of  problems:
from computer display
engineering and technologies, architecture and design to  alternative
interpretations of some observational experiments in cosmology and
unsolved problem of quantum theory \cite{cosm,penrose} (see also Conclusion).

In the first half of the present article we are suggesting concrete mathematical realization
of Poincare's ideas. Different physical geometries are being concluded from the sole geometry
of perceptive manifold and fundamental properties of physical objects --- particles, bodies,
fields. Besides quoted above Poincare's ideas on geometry's conditional role and its
dependence upon physical context, we are using a number of other ideas and observations, part
of which were expressed by some authors earlier, disregarding of our consideration.

\begin{enumerate}
%\item{\bf The role of  smooth manifolds and Lie groups.}
%As a basis for our consideration, we are taking well developed branches of analysis
%and geometry --- smooth manifolds theory, smooth mappings
%on them, and Lie groups theory that to sufficient extent reflect observed physical
%properties of space, bodies in it, and their motions.
%In practical applications we will be dealing with smooth structures
%though approach's general formulation is based upon abstract sets.
\item
{\bf Observer's role in physical theory.} This ques\-ti\-on has drawn attention mostly because of  the problems connected to
the search for quantum mechanics' due interpretation  and observables' problem
in GR \cite{grib,mitsk}.
From the very beginning we are introducing a formalized notion
 of  an observer --- his perceptive space and Newton's
mapping roughly modelling observation process together with objects of ob\-ser\-va\-ti\-on:
hyperbodies and their images in perceptive space --- ordinary bodies' histories.
It should be em\-pha\-si\-zed that observer's presence and his properties play a role
already at classical physics construction  in our approach  since
hyperbodies world's laws can have a different character than their
images in perceptive space.
\item
{\bf An essence of a physics law.}
To reveal it, we are formulating $\mathfrak{P}$-structure (physical structure) notion.
It embraces quite general understanding of a physical law:
some property observed in per\-cep\-ti\-ve space is invariant
to selection  of  some observers' subset.
It intersects as analogy in a sense of ter\-mi\-no\-lo\-gy and ideology
but doesn't coincide with Yu.I.Ku\-la\-kov's  physical structures theory \cite{kul,mich}.
\item
{\bf Physical geometry's secondary meaning.}
We are inferring $\mathfrak{P}-$geometry notion from
$\mathfrak{P}-$struc\-tu\-re
under certain assumptions. This $\mathfrak{P}-$geometry
is realized on observers' set transformation group space \cite{kok}.
\item
{\bf Normal divisor role.}
Here is an observation we make: there is a normal divisor of
Euclidean metric isometry group  ${R}^3\rtimes{O}(3)$ ---
translations' subgroup ${R}^3$, which can be identified
as physical space itself --- Euclidean space ${E}^3$ ---
by fixing one origin point.
Analogously, there is normal divisor --- translations' subgroup ${R}^4$
in isometry group ${R}^4\rtimes{O}(1,3)$ of Minkowski metric.
This subgroup is isomorphic  to  physical space of special relativity ---
Minkowski space ${M}_{1,3}.$
We are generalizing these facts and building $\mathfrak{P}-$geometry,
which has been drawn on in previous clause,
as normal divisor's inner geometry of the corresponding group if such
divisor exists. However, we have not assumed
divisor's abelian property as against given examples (see also \cite{spt}).
\end{enumerate}
We are demonstrating the approach developed here when inferring
chronogeometry.

In the second part of the article we are presenting statistically processed
data,
obtained from the experiment that's been carried out to determine
some quantitative characteristics of observer's visual space geometry.
We suggest visual manifold's affine model that satisfactorily
approximates experimental data within studied distances range.
 Achieved results are subjected to short general discussion in Conclusion.

Most of the definitions and symbols used in the first part of the article are standard.
Particularly, we denote:

$\text{Dom}(f)$ --- domain of some mapping $f$;

$\text{Im}(f)$ --- range of images of some mapping $f$;

$\text{A}\le\text{B}$ --- $A$ is subgroup of group $B$;

As new signs and definitions appear, their meaning is explained.

\section{$\mathfrak{P}$-structure and  $\mathfrak{P}-$geometry definitions}

\subsection{Properties algebra over arbitrary set}

In this section we are developing some initial foundations for
abstract properties theory. This theory being of interest in itself will be
used further on in defining $\mathfrak{P}-$structure notion.

Let's assume  $\mathcal{A}$ and $\Omega$ --- are some element sets. We will say that
{\it  $\Omega$ set  constructs properties system over $\mathcal{A}$ set}
and write it down like $\Omega=\wp(\mathcal{A}),$ if mapping $\chi:\
\mathcal{A}\times\Omega\to\mathbb{Z}_2=\{0,1\}$
is given. Element   $a\in\mathcal{A}$ will be called {\it possessing the
property}
  $\omega\in\Omega,$ if $\chi(a,\omega)=1$
and {\it non-possessing such property}   $\omega,$ if $\chi(a,\omega)=0.$

Let $\omega$ be some fixed element of the set
$\Omega.$ Designate
$\mathcal{A}_{\omega}$ and $\overline{\mathcal A}_{\omega}$ subsets of $\mathcal{A},$
satisfying the following conditions:
\[a\in\mathcal{A}_{\omega}\Leftrightarrow\chi(a,\omega)=1;\quad
b\in\overline{\mathcal{A}}_{\omega}\Leftrightarrow\chi(b,\omega)=0.
\]
Apparently
$\mathcal{A}_\omega\cup\bar{\mathcal A}_\omega=\mathcal{A}$ and
$\mathcal{A}_\omega\cap\bar{\mathcal A}_\omega=\varnothing.$ Analogously, let $a$  be some
fixed element of the set $\mathcal{A}.$ Designate
$\Omega_a$ and $\overline{\Omega}_a$ subsets of $\Omega,$
satisfying the conditions:
\[
\omega\in\Omega_a\Leftrightarrow\chi(a,\omega)=1;\quad
\delta\in\overline{\Omega}_a\Leftrightarrow\chi(a,\delta)=0.
\]
Thus, we can say that $\mathcal{A}_\omega$ ---
a set of all $\mathcal{A}$ elements possessing property  $\omega,$
$\overline{\mathcal{A}}_\omega$ --- a set of all $\mathcal{A}$ elements,
not possessing property
 $\omega,$ $\Omega_a$ --- a set of all properties possessed by an element $a$
 and  $\overline{\Omega}_a$ --- a set of all
 properties not possessed by an element $a$.

Our preceding and following description of the sets of elements and properties
over them reveals the following property    of duality:
   same mapping $\chi$
determines a set $\Omega$ as set of properties over $\mathcal{A},$ but the same mapping
determines
$\mathcal{A}$ as a set of properties on
its own properties
$\Omega.$
Therefore we can define a set $\mathcal{A}$ as $\tilde\wp(\Omega)$
--- a set of properties over $\Omega,$ where
$\tilde\chi(\omega,a)=\chi^{\text{T}}(\omega,a)\equiv\chi(a,\omega).$
Here {\it duality relations} are valid:
$$\tilde\wp(\wp(\mathcal{A}))=\mathcal{A};\ \ \wp(\tilde\wp(\Omega))=\Omega.$$
At first, in light of the above relations,  identification of the sets $\mathcal{A}$
and $\Omega$ as elements of sets and properties on them  becomes matter of agreement,
i.e. relative and conditional and we will stick to the once chosen manner.
Secondly, all definitions and relations, pertaining to one of the sets
(of elements or properties), not connected with introducing additional
structures on these sets, allow dual wording on complementary set.
We will mention it in plain form only if necessary.

Let's call property  $\omega_0\in\Omega$ {\it trivial on $\mathcal{A},$}
if
$\mathcal{A}_{\omega_0}=\mathcal{A},$ and property $\omega_\varnothing$ {\it
empty on  $\mathcal{A},$}  if
$\mathcal{A}_{\omega_{\varnothing}}=\varnothing_{\mathcal{A}}.$
Dual analogies for trivial and empty properties are {\it universal
element}
$a_0$ and  {\it transcendent element} $a_{\varnothing},$ which are identified by the relations
$\Omega_{a_{0}}=\Omega,$ and
$\Omega_{a_{\varnothing}}=\varnothing_{\Omega}.$
Note, that the existence of trivial and empty properties and their dual analogs
is not obligatory.

Since function $\chi$ puts into correspondence each element $\omega\in
\Omega$ subsets $\mathcal{A}_\omega,$ on which
boolean operations are naturally determined,
we can introduce mappings on $\Omega$, induced by boolean algebra on
$\mathcal{A}.$ Thus, let
$\top_\mathcal{A}, \bot_\mathcal{A}, \varrho_\mathcal{A} $  --- correspondingly, binary
operation, unary operation, and
binary relation on $\mathfrak{B}(\mathcal{A}),$
where $\mathfrak{B}(\mathcal{A})$ --- set of all subsets of $\mathcal{A},$
called  {\it boolean.} Then it is possible to determine the corresponding
one- and two- component functions
   and binary relation:
$\top_\Omega:\ \Omega\times\Omega\to\Omega'',\ \bot_\Omega:\ \Omega\to\Omega',\
\varrho_\Omega$ according to following rules:
\[ \omega_1\top_{\Omega}\,\omega_2=\omega_3\Leftrightarrow
\mathcal{A}_{\omega_3}=\mathcal{A}_{\omega_1}\top_\mathcal{A}\,\mathcal{A}_{\omega_2},\
\text{for all}\ \omega_1,\omega_2\in\Omega; \]
\[ \bot_{\Omega}\omega_1=\omega_2\Leftrightarrow
\mathcal{A}_{\omega_2}=\bot_{\mathcal{A}}\mathcal{A}_{\omega_1},\ \quad\text{for all}\quad
\omega_1\in\Omega;\]
\[\omega_1\varrho_{\Omega}\omega_2\Leftrightarrow
\mathcal{A}_{\omega_1}\varrho_{\mathcal{A}}\mathcal{A}_{\omega_2}\ \quad\text{for all}\quad
\omega_1,\omega_2\in\Omega.\]
 Here $\Omega''\supseteq\Omega,\
\Omega'\supseteq\Omega$ --- some new sets of properties, which can be produced from
 the initial set $\Omega$  by com\-ple\-ti\-on over corresponding one or two component
 mappings.

Let's designate $\Omega_\infty$  completion of  $\Omega$ by whole
boolean
algebra, generated by elements $\{\mathcal{A}_{\omega}\}_{\omega\in\Omega}$
and let's call it {\it complete properties system over $\mathcal{A}$, induced by $\Omega.$}
Initial set $\Omega$ will be called {\it generating} for $\Omega_\infty.$
Note, that set
$\mathcal{A}$ becomes
    topological space with topology
$\{\mathcal{A}_{\omega'}\}_{\omega'\in\Omega_\infty}$ with  prebase
$\{\mathcal{A_{\omega}}\}_{\omega\in\Omega}.$

We are going to introduce the following cor\-res\-pon\-den\-ce rules between boolean operations
   designations in $\mathcal{A}$ and the corresponding operations and relations in
$\Omega_\infty:$
\begin{equation}\label{prop}
\mathcal{A}\to\mathbf{1};\ \varnothing_{\mathcal{A}}\to\mathbf{0};\ \cup\to+;\
\cap\to\cdot;\
\subseteq\to\ge;\ \bar{}\to\mathbf{1}-;\ \setminus \to
 -; =\to=.
\end{equation}

The properties, connected by a relation "$=$"\,
 will be called {\it equivalent} on  $\mathcal{A},$
 and  if $\omega_1\ge\omega_2,$ we'll say, that  $\omega_1$  {\it is
 not weaker than $\omega_2$} on $\mathcal{A}.$
  It is easy to see that the introduced earlier set $\bar
{\mathcal A}_\omega\equiv\mathcal{A}_{\mathbf{1}-\omega}.$
 Obviously, the properties similar to those of boolean
 operations and relations  in $\mathcal{A}$ are satisfied by all introduced operations and
   ratios between elements of  $\Omega_\infty$
(commutativity, associativity, distributivity, etc).
For instance, it easy to check the following identities of the obtained properties algebra:
\[
(\omega_1\pm\omega_2)\cdot\omega_3=\omega_1\cdot\omega_3\pm\omega_2\cdot\omega_3;\
\omega^2=\omega;\
(\mathbf{1}-\omega)\cdot\omega=\mathbf{0}.
\]
Note, that in $\Omega_\infty$ $\omega_0=\mathbf{1}$, $\omega_{\varnothing}=\mathbf{0}.$

Concluding this paragraph, let's introduce a notion  of
{\it determinable elements set $\mathcal{D}\subset\mathcal{A},$}
whose  characteristic is that  there exists (perhaps not the unique!)
family of properties
$\{\omega_i\}_{i=1,\dots, N}\subset\Omega,\  N<\infty $, that
$\mathcal{D}=\mathcal{A}_{f(\omega_1,\dots,\omega_N)},$
where $f(\omega_1,\dots,\omega_N)$ --- finite superposition of (\ref{prop})
with $\{\omega_i\}_{i=1,\dots, N}.$
In other words, determinable elements set $\mathcal{D}$  can be determined
     (described) using finite number of properties from  $\Omega.$
     By properties set definition, some sets from $\mathcal{A}$
     are always determinable,
      therefore any $\mathcal{A}$ is {\it partially determinable.}
      Accordingly, let's call $\mathcal{A}$ {\it quite determinable,}
      if a system of de\-ter\-mi\-na\-ble sets
$\{\mathcal{A}_\alpha\}$ coincides with
      boolean $\mathfrak{B}(\mathcal{A}).$

\subsection{Newton's mapping and hyperclasses.}

The next point of our $\mathfrak{P}$-structure  defining will be
two sets: {\it world set} $\mathcal{M}$ and {\it perceptive set of
events}
$\mathcal{N}$.
The former  is going to contain elements of  "true"\, physical world,
the latter ---  elements of its perception by some observer.

Since space and time perceptions  exist to a significant extent
independently,
we assume that perceptive set of events $\mathcal{N}$ is direct product $\mathcal{T}\times\mathcal{V},$
where $\mathcal{T}$ --- {\it time perceptive set}, $\mathcal{V}$  ---
{\it visual perceptive set,}
whose connection with physical time and space is to be defined later on.
Set $\mathcal{V}$ is in bijective correspondence with
{\it simultaneous event space} $\mathcal{V}_T,$
that we define as section $\{T\}\times\mathcal{V}$ of set $\mathcal{N}$
by some element of time set $T\in\mathcal{T}$.
The sets $\mathcal{T}$ and $\mathcal{V}$ will be considered to
be metrized with metrics
$\tau$: $\mathcal{T}\times\mathcal{T}\rightarrow R$
and $\eta$:
$\mathcal{V}\times\mathcal{V}\rightarrow R.$

Assume further that $\mathcal{N}$ is an image of some surjective
mapping $f$: $\mathcal{M}\rightarrow\mathcal{N}$,
which, in physical language, describes a process of observation of some subset
of world set made by some observer. We are going to call this mapping
{\it Newton's one.}

Let's define {\it hyperbody} $\mathcal{B}$
as some subset of a set $\text{Dom}(f)\subseteq\mathcal{M}.$ Its image
$f(\mathcal{B})\subseteq\mathcal{T}_{\mathcal{B}}\times B,$ where
$\mathcal{T}_\mathcal{B}=(\pi_1\circ f)(\mathcal{B})\equiv
f_{\mathcal{T}}(\mathcal{B}),$ $B=(\pi_2\circ f)(\mathcal{B})\equiv
f_{\mathcal{V}}(\mathcal{B})\equiv\cup_{T\in\mathcal{T}}B_{T},$
$\pi_1,\pi_2$ --- projections
 $\mathcal{N}\rightarrow\mathcal{T},$
$\mathcal{N}\rightarrow\mathcal{V}$ correspondingly and a
designation $B_T$ for instant body at a time point $T$: $B_T\equiv\mathcal{V}_T\cap
f(\mathcal{B})$  is introduced.
This image can be understood as some subset of
direct product $\mathcal{T}\times$ $\mathfrak{B}(\mathcal{V}),$
which, in turn, defines mapping $\mathcal{T}\ni T\mapsto
B_T\in\mathfrak{B}(\mathcal{V}).$
A graph of this mapping will be referred to as {\it $f$-history of body $B$}
induced by a hyperbody $\mathcal{B}.$

Using mapping $f$ on the set of all hyperbodies
in $\mathcal{M}$, we can introduce canonical equivalency:
two hyperbodies $\mathcal{B}_{\alpha}$ and $\mathcal{B}_{\beta}$ are {\it
$f$-equivalent,}
if $f(\mathcal{B}_{\alpha})=f(\mathcal{B}_{\beta}).$
Equivalency class $[\mathcal{B}]_f$ of some hyperbody $\mathcal{B}$
will be called its {\it $f$-hyperclass.}
Particularly, elementary event  $P\in
\mathcal{N}$ preimage is some hyperbody $\mathcal{M}_P,$
thus $\mathcal{M}'=\text{Dom}(f)/\mathcal{M}_P$
is canonically isomorphic $\mathcal{N}.$
It is also possible to introduce the sets $\mathcal{M}_T\equiv f^{-1}(\mathcal{V}_T)$ and
$\mathcal{M}_V\equiv f^{-1}(\mathcal{T}_V),$ where $\mathcal{T}_V\equiv\mathcal{T}\times\{V\}$ ---
{\it the history of a point $V\in\mathcal{V}.$} If $P=(T,V)\in\mathcal{N},$
than it's obvious that $\mathcal{M}_P=\mathcal{M}_T\cap\mathcal{M}_V.$

Thus, any member of $f$-hyperclass gives identical $f$-history
in $\mathcal{N}$, that, in turn, can be understood as "movement
graphics"\,
of some body $B$ in $\mathcal{V}$, perceived by an observer.
Alongside with hyperbodies and hyper\-clas\-ses, we can consider as well their
unions $\cup_{\alpha}\mathcal{B}_{\alpha}$ and their corresponding $f$-hyperclasses,
which will generally spe\-a\-king be transformed by Newton's mapping
into compound history of a bodies system $B_{\alpha}$:
$f(\cup_{\alpha}\mathcal{B}_{\alpha})\subseteq\cup_{\alpha}(\mathcal{T}_{\mathcal{B}_{\alpha}}\times
B_{\alpha}).$
It is natural to call these hyperbodies {\it compound.}
Later on, if it's not specially stipulated, it will be dealt only with hyperclasses,
omitting  square brackets, where it is appropriate and will not lead to a mess.

 Note, that using  $\tau$ and $\eta$ and  mapping $f,$
 it's possible to introduce functions $\tau^{\ast}$ and $\eta^{\ast}:\
\text{Dom}(f)\times\text{Dom}(f)\rightarrow R,$
%\begin{equation}\label{metr}
\[\tau^{\ast}(p,q)\equiv\tau(f_{\mathcal{T}}(p),f_{\mathcal{T}}(q));\quad
\eta^{\ast}(p,q)\equiv\eta(f_{\mathcal{V}}(p),f_{\mathcal{V}}(q))
\]
for
all $p,q\in\text{Dom}(f),$ which can be considered as metrics on factor-set  $\mathcal{M}'.$

\subsection{Newton's mappings transformation group}

Up to now we have dealt with only one observer, to be more exact --- with his Newton's mapping.
The existence of observers' set as well as a possibility of one observer's state change
(for example his motion) will be highlighted in our construction by means of
some subgroup ${\mathcal{G}}$ of general  group of automorphisms $\text{Aut}(\mathcal{N}).$
We are not specifying this subgroup at this stage but  major role in applications will be
played by finite-dimensional Lie groups acting on $\mathcal{N}.$

In fact, the very group character of mappings  from ${{ \mathcal{G}}}$ enables to make a
number of important conclusions regarding behaviour  of the objects introduced in
previous paragraph  with ${{
\mathcal{G}}}$ acting.

\medskip
{\bf The Proposition 2.1} {\tt
Automorphisms $g\in\mathcal{G}$ indu\-ce transformations of
Newton's mapping:}
\[f\rightarrow f_g=g\circ
f.\]
We are postulating that {\it  any automorphism $g$ from ${{ \mathcal{G}}}$,
acting on some Newton's mapping, results again in Newton's mapping}.
Thus, we can say that {\it observers family
is a homogeneous space with respect to a group ${{\mathcal{G}}}$ action.}
Let's designate this family $\mathcal{O}_f\equiv\{f_{g}\left|\, {g}\in\mathcal{G}\right.\}$.
Any two families $\mathcal{O}_f$ and $\mathcal{O}_{f'}$ either
coinside or not intersect. More exactly , relation of $\mathcal{O}_f$
and $\mathcal{O}_{f'}$ is ruled by

\medskip
{\bf The Proposition 2.2} {\tt Any $\mathcal{O}_f$ and $\mathcal{O}_{f'}$
coinside  only if $\text{zer}(f)=\text{zer}(f'),$ where $\text{zer}$
- fibers de\-com\-po\-si\-ti\-on of domain of some mapping.}
\smallskip

{\bf Proof.} Let $\mathcal{O}_{f'}=\mathcal{O}_{f},$ then
$f'=g\circ f$ for some $g\in\mathcal{G}.$ Going to its factors,
we have:
\begin{equation}\label{fact}
\text{fact}\, f'= \text{fact}(g\circ f)=g\circ\text{fact}\,f,
\end{equation}
since $g$ --- automorphism. Here $\text{fact}\,f$: $\mathcal{M}/\text{zer}(f)=
\mathcal{M}'\to\mathcal{N}.$ It implies $\text{zer}(f)=\text{zer}(f'),$
since $\text{zer}=\text{Dom}(\text{fact}).$ Inversely, if $\text{zer}(f)=\text{zer}(f')$,
then there exists $g\in\text{Aut}(\mathcal{N}),$ such that (\ref{fact})
takes place. However, $g$ can be lying in $\text{Aut}(\mathcal{N})\setminus\mathcal{G},$
so condition of the statement is only necessary, but not
sufficient.$\Box$

\smallskip
\noindent  As a consequence we have

\medskip
{\bf The Proposition 2.3} {\tt
The set $\mathcal{M}'$ is invariant under ${{ \mathcal{G}}}$ action.}

\smallskip
{\bf Proof.} It follows from the isomorphism $\mathcal{M}'\sim\mathcal{N}$
 that $g\in\mathcal{G}$ induces automorphism $g_\ast:\
\mathcal{M}'\to\mathcal{M}',$
where $g_\ast=(\text{fact}\,f)^{-1}\circ g\circ \text{fact}\,f.\,\Box$

Let
$g(\mathcal{N})=\mathcal{N}'=\mathcal{T}'\times\mathcal{V}'.$
We will be saying  that  $g$ is automorphism of
$\mathcal{T}$-type ($g\in {{ \mathcal{G}}}_{\mathcal{T}}$),
if for any $T\in\mathcal{T}$  there exists the only $\
T'\in\mathcal{T}',$ such as
$g(T,\pi_2\mathcal{V}_T)=(T',\pi_2\mathcal{V}'_{T'}),$ i.e.
$g$ defines bijection between the sets of simultaneous
events in $\mathcal{N}$ and $\mathcal{N}'.$
Similarly, we will call $g$  automorphism of $\mathcal{V}$-type
$(g\in {{ \mathcal{G}}}_{\mathcal{V}})$,
if for any $
V\in\mathcal{V}$  there can be found the only $V'\in\mathcal{V}',$ such
as $g(\pi_1\mathcal{T}_V,V)=(\pi_1\mathcal{T}'_{V'},V'),$
i.e.  $g$
defines bijection  between histories of
point-events in $\mathcal{N}$ and $\mathcal{N}'.$ And finally, we'll call $g$
automorphism of $\mathcal{T}\mathcal{V}$-type ($g\in {{
\mathcal{G}}}_{\mathcal{T}\mathcal{V}}$) if at the same time $g\in {{
\mathcal{G}}}_{\mathcal{T}}$ and  $g\in {{ \mathcal{G}}}_{\mathcal{V}}.$
It is easy to show that the above described bijection  can be
plainly written down as follows:

\[g_{\mathcal{T}}=\left\{\begin{array}{lcl} T'&=&T'(T);\\
V'&=&V'(T,V),\end{array}\right.g_{\mathcal{V}}=\left\{\begin{array}{lcl} T'&=&T'(T,V);\\
V'&=&V'(V),\end{array}\right.
\]
\[
g_{\mathcal{TV}}=\left\{\begin{array}{lcl} T'&=&T'(T);\\
V'&=&V'(V),\end{array}\right..\]
%where $T'(\cdot),V'(T,\cdot),$\ $T'(\cdot,V),V'(\cdot),$\
%$T'(\cdot),V'(\cdot)$ --- are correspondent bijections.
This  representation obviously shows that each type of  automorphisms
form subgroup of ${{ \mathcal{G}}}$ (see also \cite{simult}).

\subsection{Sets $\text{Cont}(\mathcal{N}),$  ${M}_D,$ $\mathfrak{P}$-structures
and $\mathfrak{P}$-geometries.}

Let us introduce a set  $\text{Cont}(\mathcal{N})$ in the category
$\text{SETS}$ as a maximal subset of morphisms epicocones beginnings\footnote{
Lets remind that for any  category   $\mathfrak{R}$
morphisms cocone  $\text{C}^\ast\,(A)$
with vertex $A\in\text{Ob}\,\mathfrak{R}$ is a system  of morphisms
$\alpha_i:\,A_i\to A,$ $i\in I.$ Cocone  is called dense
or epicocone, if from the
$\varphi\circ\alpha_i=\psi\circ\alpha_i$ for all  $i\in I$ it follows
$\varphi=\psi.$ Here "$\circ$"\, means standard (from right to left)
composition of morphisms. In other words,
epicocone  $\text{EpiC}^\ast(A)$ consist of all epimorphisms
$\text{Epi}\,\mathfrak{R}$ of the category $\mathfrak{R}$ with end $A.$
Within the sets category $\text{SETS}$ epimorphisms coinside with surjections.
The notions of  cocone and epicocone are dual to cone and monocone
ones within the given category $\mathfrak{R}$ \cite{algebra}.}
$\text{EpiC}^\ast(\mathcal{N})\subset\text{Mor}(\mathfrak{A},\mathcal{N}),$
where $\mathfrak{A}\in\text{Ob}\,\text{SETS},$ each element
$\mathfrak{A}$ of which:

a) is connected with  $\mathcal{N}$
by  (perhaps nonunique) epimorphism
$\sigma\in\text{EpiC}^\ast(\mathcal{N})$;

b) is
{\it $(\mathcal{T},\mathcal{V},R)$-autonomous,} i.e.
can be  defined by some universal symbolic formula of the following kind:
\begin{equation}\label{form}
\mathfrak{A}=\Lambda(\mathcal{T},\mathcal{V},R),
\end{equation}
where $\Lambda=\alpha_1\circ\cdots\circ\alpha_N$ $(N<\infty)$ and every
morphism $\alpha_i$ is either {\it universal}, or {\it $R$-morphism} of the category
SETS. Morphism $\alpha$ of  $\text{SETS}$ we'll call  universal,
if its definition don't refer to any additional structures  on the sets
of the class $\text{Ob}\,\text{SETS}.$
Morphism $\alpha$ we'll call  $R$-morphism,
if  $\text{Im}\,\alpha\subseteq R^n$ or $\text{Dom}\,\alpha\subseteq R^n$
under some $n<\infty.$
Futhermore for the brevity sake we'll write  (\ref{form})
in the compact form  $\mathfrak{A}=\Lambda(\mathcal{N}).$

Let us make some comments to the above given definitions. Laws of
nature, which we open analyzing our experiments and working
out theories,  may be rather complicated to be formulated  in terms
of primary observers perceptions, i.e. in terms of
$\mathcal{N},\mathcal{T}$ and $\mathcal{V}.$
For example, if we temporarily assume
$\mathcal{T}=E^1,$ $\mathcal{V}=E^3,$  then Lagrange
mechanics of  $n$ matter points is formulated in terms
of tangent bundle $\top (E^3)^{\times n},$ while Hamiltonian mechanics
 --- in terms of cotangent bundle $\top{}^\ast(E^3)^{\times n}$
 (Lagrange and Hamilton functions are defined on this
manifolds). Solid dynamics in classical mechanics can be described
by means of Lagrange function, defined on manifold
$\top E_3\times \top \text{SO}(3),$ where $\text{SO}(3)$
--- proper orthogonal group of  $E_3$,
isomorphic to a space of solids angle positions.
Classical ele\-ctro\-dy\-na\-mics demands con\-si\-de\-ra\-ti\-on of vector and tensor
bundles over Minkowski space $M_{1,3}$. Abstract set $\text{Cont}(\mathcal{N}),$
which has been defined above, is wide (in some sense maximal)
arena for formulation of physical laws with any degree of complexity.
In our examples Cartesian product  "$\times$"\,, is example of
universal morphism
$A\to A\times B,$
while tangentialization $\top$ and cotangentialization
$\top^\ast$ ope\-ra\-ti\-ons are important particular cases of nonuniversal
$R$-mor\-phisms of the category SETS, in fact, acting in  category
$\text{DIFF}$ of smooth manifolds, which is subcategory of $\text{SETS}.$
By means of coordinate homeomorphisms  the mapping  $\top $ and $\top^\ast$
always can be realized as family of embeddings
$R^n\to R^{2n}$ for some $n.$

%Free generators of this algebra are images
%of the mappings that are in turn
%different finite degrees (compositions) of
%{\it tangelization} operation $T:\ \mathcal{Q}\to T\mathcal{Q},$
%{\it cotangelization} operation $T^\ast:\ \mathcal{Q}\to
%T^\ast\mathcal{Q}$ and {\it booleanization} operation
%$\mathfrak{B}:\ \mathcal{C}\to \mathfrak{B}(\mathcal{C})$
%with domain $\mathcal{N}.$ Here  $\mathcal{Q}$ is some free
%generator of of algebra ${Cont}(\mathcal{N})$,
%which is smooth manifold of finite dimension, $\mathcal{C}$
%---  arbitrary free generator of algebra ${Cont}(\mathcal{N})$.
%A typical element of ${Cont}(\mathcal{N})$ is an expression:
%\[\mathcal{N}^{\times
%n}\times\mathfrak{B}(\mathcal{Q})^{\times
%b_1}\times\dots\times\mathfrak{B}^k(\mathcal{Q})^{\times
%b_k}\times\dots
%\]
%\[\dots\times(T^l(\mathcal{N}))^{\times t_l}\times\dots\times(T^{\ast s}(\mathcal{N}))^{c_s},\]
%where $n,b_i,l,t_l,s,c_s,\dots$ --- natural numbers and powers of operations
%$T,\ T^\ast$ and $\mathfrak{B}$  are understood as their
%appropriate iterations. From physical point of view, the space
%${Cont}(\mathcal{N})$  is quite broad and abstract arena for
%formulating physical laws of almost any degree of difficulty in a language of perceptive manifold
%elements\footnote{Starting from this paragraph, we assume, using (co)tangelization operations,
%that $\mathcal{N}$ has a structure of smooth manifold.
%This is quite natural from a viewpoint  of "smoothness"\,
%of the overwhelming majority of physical problems.}.

Our construction, of course, is based on some par\-ti\-cu\-lar role of the set
of real number $R$ within class of objects of the category SETS.
Cause of this particularity is clarified by the following definition.
For any  $\Lambda$ from (\ref{form}) and every
natural $D$ lets define the set:
\begin{equation}\label{mapp}
\text{M}_{D}(\Lambda)\equiv\text{Maps}(\text{Im}(\Lambda),R^D)\equiv\{f:\
\mathfrak{A}\rightarrow R^D\,|\,\mathfrak{A}=\Lambda(N)\}
\end{equation}
of all mappings of the sets $\Lambda(\mathcal{N})$ in $R^D$ with fixed  $\Lambda$
and $D.$
Let designate
\[
\beth_{\mathcal{N}}\equiv \bigcup\limits_{D=0}^\infty\bigcup\limits_{\Lambda}\text{M}_{D}(\Lambda).
\]
The set $\beth_{\mathcal{N}}$  is, in physical context,
the collections of various "arithmetizations"\, and  "metrics"\, that enable us to
fix measuring devices data
and measurements' results in the form of number tables, functions, graphs,
correlations  and to derive quantitative (value) results from the abstract theories,
formulated in terms of some sets  from $\text{Cont}(\mathcal{N})$.

%For any element $\mathcal{L}\in{Cont}(\mathcal{N})$ and each natural
%$D$ let us introduce a set:
%\begin{equation}\label{mapp}
%{M}_{D}(\mathcal{L})\equiv{Maps}(\mathcal{L},\mathbb{R}^D)\equiv\{f:\
%\mathcal{L}\rightarrow\mathbb{R}^D\}
%\end{equation}
%of all mappings from $\mathcal{L}$ to $\mathbb{R}^D.$
%Let us designate
%\[
%\aleph=\{{M}_{D}(\mathcal{L}),\ \mathcal{L}\in{Cont}(\mathcal{N}),\, D\in \mathbb{N}\}.
%\]
%The set $\aleph$  is , in physical context,
%the collections of various "arithmetizations"\, and  "metrics"\, that enable us to
%fix measuring devices data
%and measurements' results in the form of number tables, functions, graphs, correlations  etc.

Let's define homomorphism
\[\Delta_{(\Lambda,D)}:\
{\mathcal{G}}\rightarrow\text{Aut}(\text{M}_D(\Lambda)),
\]
designed as follows. Let $\mathfrak{A}=\Lambda(\mathcal{N}),$
where $\Lambda$ --- function of type (\ref{form}),
defining $\mathfrak{A}$ by means of some finite set of suitable morphisms.
If under
${g}\in\mathcal{G},$ $\mathcal{N}\to \mathcal{N}'={g}(\mathcal{N}),$
then homomorphism $\chi_g:\ \mathfrak{A}\to
\mathfrak{A}'=\Lambda(\mathcal{N}'),$
is defined, such as for diagram:
\[
\begin{CD}
\mathfrak{A}@>\chi_g>>\mathfrak{A}'\\
@A\Lambda AA@AA\Lambda A\\
\mathcal{N} @>g>>\mathcal{N}'
\end{CD}
\]
commutativity  condition is met:
$\Lambda\circ g=\chi_g\circ\Lambda.$
We define then
$\Delta_{(\Lambda,D)}(g)\varphi\equiv\varphi\circ \chi_g$ for any
$\varphi\in\mathrm{M}_D(\Lambda).$

Let's designate the set
$\mathfrak{A}_{\mathcal{B}}\equiv(\Lambda\circ f)(\mathcal{B})\equiv\Lambda_{\mathcal{B}}(\mathcal{N}),$
consisting of those
elements of $\mathfrak{A},$  whose preimage under mapping
$\Lambda$ were lying in $f(\mathcal{B})\subseteq\mathcal{N},$ where
$\mathcal{B}$ is some hyperbody from   $\mathcal{M},$ and $f$ ---
Newton's mapping of some observer.
 Consider as well the set of  properties $\wp(M_D(\Lambda))$ over
$\text{M}_{D}(\Lambda).$ Now we can formulate the notion of
$\mathfrak{P}$-structure.

%\subsection{$\mathfrak{P}$-structure definition}

\noindent Lets call the collection
$\langle\{\mathcal{B}^{\omega}\},f,
\mathcal{G}^{\omega},\Lambda,\text{M}_D(\Lambda_{\omega}),\omega
\rangle,$  {\it $\mathfrak{P}$-structure induced by a property $\omega$}
or, in short, {\it $\mathfrak{P}^{\omega}$-structure,}
if
\begin{equation}\label{Pstr}
\text{id}_{\text{M}_D(\Lambda)}\neq\Delta_{(\Lambda,D)}(\mathcal{G}^\omega)\le
\text{Aut}(\text{M}_D(\Lambda)).
\end{equation}
%for any $\mathcal{B}^\omega\in\{\mathcal{B}^\omega\}.$
In the above mentioned collection:

$\{\mathcal{B}^{\omega}\}$ ---
 collection of some hyperbodies  from $\mathcal{M}$;

$f$ ---  some Newton's mapping;

$\mathcal{G}^{\omega}$ --- some subgroup of the group
$\mathcal{G}\in\text{Aut}(\mathcal{N})$;

$\Lambda$ --- some function from
$\mathcal{N}$ in  $\text{Cont}(\mathcal{N});$

$\text{M}_D(\Lambda_{\omega})$ ---  
subset of mappings from ${M}_{D}(\Lambda)$,
acting from $\mathfrak{A}_{\mathcal{B}^{\omega}}$
into $R^D$ and having a property $\omega$;

$\omega$ --- an element from
$\wp(M_D(\Lambda)).$

In other words, $\mathfrak{P}$-structure is formed every time when for any member of
the family of hyperbodies
$\{\mathcal{B}^\omega\}$  the property $\omega$ over
${M}_D(\Lambda)$ remains invariant with respect to some nontrivial
subgroup of automorphisms
$\Delta_{(\Lambda,D)}(\mathcal{G}^\omega)\le
\mathrm{Aut}(\mathrm{M}_{D}(\Lambda)).$

%\subsection{$\mathfrak{P}$-geometry definition}

We are going to say that $\mathfrak{P}^\omega$-structure induces {\it
$\mathfrak{P}^\omega$-geometry,} if $ \mathcal{G}^{\omega}$:

1) has a structure of semidirect product:
$\mathcal{G}^\omega={N}^{\omega}\rtimes{S}^{\omega},$
where ${N}^{\omega}\trianglelefteq
\mathcal{G}^{\omega}$ --- normal divisor,
${S}^{\omega}\leq\mathcal{G}^{\omega}.$

2) is an isometry of some metric $\rho^{\omega}$:
${N}^{\omega}\times{N}^{\omega}\rightarrow R$, defined on
${N}^{\omega}$, i.e.
\begin{equation}\label{isometry}
\rho^{\omega}({gg}_1,{gg}_2)=\rho^{\omega}({g}_1,{g}_2),
\end{equation}
for any $ {g}_1, {g}_2\,\in\, {N}^{\omega}, {g}\in
\mathcal{G}^{\omega}.$

Let us clarify the implication of the introduced notions and definitions by means of the
following chain scheme, connecting objects of the outer physical world on the left side and our
geometrical  notions about them:

{\it hyperclasses $[\mathcal{B}]$} $\stackrel{\scriptstyle f}{\rightarrow}$ {\it
 the set of observed histories}
$\stackrel{\scriptstyle\mathfrak{P}^\omega,\mathcal{G}^{\omega}}{\rightarrow}$
$\mathfrak{P}^\omega$-{\it structure}
$\stackrel{\scriptstyle\rho^{\omega}}{\rightarrow}$ $\mathfrak{P}^\omega$-{\it
geometry}

First arrow plainly introduces an observer into the scope of physical consideration, or to be more
exact, his apparatus of surrounding world perception, ma\-the\-ma\-ti\-cal\-ly formalized via Newton's
mapping. It can turn out that the answer to the question about specific type of this
mapping doesn't lie entirely within the frames of a certain physical theory or physical experiment but
should be based as well upon the data of other branches of science: neurobiology, perception
psychophysiology, and, perhaps,  psychology (see, for example
\cite{krylov,penrose}).
   Second arrow shows that the most general (abstract) physical structures,  which contain,
among other things, the ca\-te\-go\-ry called laws of physics, are formed under two necessary conditions:
1) hyperclasses are not arbitrary, but struc\-tu\-red in a certain way in
$\mathcal{M};$
2) this struc\-tu\-red\-ness possesses a certain (and non-trivial) steadiness with respect to the change of
one observer with the other.
First condition can in plain language be attributed to the properties of objective reality of the outer
world, the second --- to the identity of perception apparatus of different observers and to the notion of
invariance, which, in one way or the other, accompany any physical law.
The third arrow is closely connected to Poincare's ideas that have been touched upon in the
Introduction: geometry, within which we are constructing our models of physical reality, has two
foundations in itself: outer --- physical properties of the objects we are dealing with in theory and
practice, and inner --- non-empty intersection of the groups of automorphisms
of perceptive manifold
and symmetry groups of the corresponding $\mathfrak{P}$-structure.
The appropriate $\mathfrak{P}$-metric, i.e. physical
geometry, ap\-pe\-ars as a notion derived from the observed properties of some hyperclasses that are at
first formulated on the set $\text{Cont}(\mathcal{N})$ and the
properties (topological, metrical, etc.) of perceptive space
$\mathcal{N}.$

\section{Group space metrization}\label{teor}

We remind that a non-negative function $\rho:\
\mathcal{A}\times\mathcal{A}\rightarrow{R},$ meeting the following characteristics:
\begin{enumerate}
\item
$\rho(a,a)=0,\ \quad\text{for all}\quad a\in \mathcal{A};$
\item
$\rho(a,b)=\rho(b,a),$ for all $a,b\in\mathcal{A};$
\item
$\rho(a,b)\le\rho(a,c)+ \rho(c,b),$ for all $a,b,c\in\mathcal{A}$
\end{enumerate}
is called metric on an arbitrary set $\mathcal{A}.$
In our case, the role of a set $\mathcal{A}$ is played by group space of normal divisor
${N}^{\omega}$ of group $\mathcal{G}^{\omega}.$
%Since any Lie group is a manifold, then it is at the same time a regular space
%satisfying the second
%countability axiom, and thus, due to Uryson's theorem, a metrizable space \cite{top}.

At the first stage of considering various
$\mathfrak{P}^\omega$-struc\-tu\-res, there appears the group
$\mathcal{G}^{\omega},$ which is a solution to property invariance equations  (\ref{Pstr}).
Further, in case this group is split into
normal divisor ${N}^{\omega}$ and the group of its external automorphisms
${S}^\omega,$ metric is found from the
condition of isometry
(\ref{isometry}), which supplements conditions 1, 2, 3, defining metric.
To solve the
appropriate func\-ti\-o\-nal equations using analysis method, it is necessary to
demand enough smoothness
for function $\rho.$ Everywhere below we will assume $\rho\in C^k\times C^{k},\ k\ge2.$

 Let us formulate a number of propositions enabling to construct the families of metrics given that
some initial function with special properties is known. Namely, assume that function $\rho_0$,
meeting
isometry condition (\ref{isometry}), is found.

\medskip
{\bf The Proposition 3.1}  \tt  In case function $\rho_0$ meets \it metricity
condition:
\begin{equation}\label{crit}
|\rho_0(c,c)|\le|\rho_0(a,c)|+|\rho_0(c,b)|-|\rho_0(a,b)|,\
\end{equation}
\tt for all $a,b,c\in\mathcal{A},$
then function
\begin{equation}\label{metr}
\rho(a,b)\equiv C [|\rho_0(a,b)|+|\rho_0(b,a)|-|\rho_0(a,a)|-|\rho_0(b,b)|],
\end{equation}
$(C\in R_+)$ meets conditions 1, 2, 3 and, the\-re\-fo\-re, is a metric, satisfying isometry
con\-di\-ti\-on
(\ref{isometry}).

\smallskip
\noindent\rm Function $\rho_0$ will further be referred to as {\it generating,} and generating function,
satisfying theorem's
condition  {\it metric.}

\medskip
{\bf The Proposition 3.2} \tt If smooth function $\varphi$: ${R}_+\rightarrow{R}_+$ of
class $C^k,\ (k\ge2)$,
satisfying
the following requirements:
\begin{enumerate}\label{prop1}
\item
$\varphi(0)=0;$
\item
$\varphi$ - isotonic, i.e. $\varphi(x)\le\varphi(y),$ always, when  $x\le
y,$ for all $x,y\in R_+;$
\item
$\varphi$ - {\it  lowering} i.e.
$\varphi(x+y)\le\varphi(x)+\varphi(y),$
for all $x,y\in{R}_+$
\end{enumerate}
is given, then, if $\rho$ is metric on $\mathcal{A},$ so
 $\rho_{\varphi}\equiv\varphi\circ\rho$ -
 metric on $\mathcal{A}$ as well.

\medskip
{\bf The Proposition 3.3} \tt Let smooth function $\Phi:$ ${R}^n_+\rightarrow{R}_+$
of class $C^k,\ k\ge2$   satisfies the fol\-lo\-wing requirements:
\begin{enumerate}
\item
$\Phi(0,\dots,0)=0;$
\item
$\Phi$  - isotonic, i.e.
\[\Phi(x_1,\dots,x_n)\le\Phi(y_1,\dots,y_n),\]
always, when $x_i\le y_i,$ for all $i=1,\dots,n;$
\item
$\Phi$ -  lowering, i.e.
\[\Phi(x_1+y_1,\dots,x_n+y_n)\le\Phi(x_1,\dots,x_n)+\Phi(y_1,\dots,y_n),\]
for all $\{x_i\}, \{y_i\}\in{R}_+^n.$
\end{enumerate}
Further, let $\{\rho_i\}_{i=1,\dots,n}$ - a set of metrics on $\mathcal{A}$,
meeting\ \ \  isometry\ \  condition \ \  (\ref{isometry}).\ \ Then \newline
$\Phi(\rho_1,\dots,\rho_n)$ -
metric on $\mathcal{A},$
satisfying iso\-met\-ry condition.

\smallskip
\noindent\rm   Propositions 3.1-3.3 are proved by straight check-up on satisfiability of conditions of
metric 1,2,3.

\medskip
{\bf The Proposition 3.4} \tt  Assume that function $\Phi_0:$
${R}^n\rightarrow{R}$ meets the conditions:
\begin{enumerate}
\item
$\Phi_0\in C^k(\mathbb{R}_+^n),\ k\ge2;$
\item
$\displaystyle \frac{\partial\Phi_0}{\partial x_i}>0,\quad\text{for
all}\quad
\{x_i\}\in\mathbb{R}^n_+;$
\item
Quadratic form $\partial^2\Phi_0$
is negatively defined on $T{R}^n_+\times T{R}^n_+.$
\end{enumerate}
Then function $\Phi\equiv\left(\Phi_0-\Phi_0(0)\right)\left|_{{R}_+^n}\right.$
satisfies the conditions of the theorem 5.3.

\noindent \rm The idea of  proving holds that with
the conditions of the theorem satisfied, function $\Phi$ graph about any point $P$
appears like a piece of
"convex up"\, surface, for which it is straightly checked if the condition of
proposition 3.3 holds true locally. Global validity of the theorem follows from
transitive property of the relation "$\le.$".

\section{Example: Physical Time of the classical mechanics}
\subsection{$\mathfrak{P}$-structure of physical time in classical mechanics}

$\mathfrak{P}$-structure of classical time is based upon three fun\-da\-men\-tal facts,
which are directly detected by an observer
$f$ in $\mathcal{N}$: the existence of {\it elementary bodies,} the existence of {\it physical
patterns of time} and {\it time arrow.} Let us introduce the appropriate mathematical
wording\footnote{Hereafter we assume that $\mathcal{N}$ has a structure of 4-dimensional
differentiable manifold. It means, in particular, that we can use, if necessary, local real
coordinates. Also, we assume, if not otherwise specified, that $\mathcal{G}$ --- Lie group.}.

1) A set $\mathcal{E}\subset\mathfrak{B}(\mathcal{M}),$ any element $\mathcal{B}^{\text{e}}$
of which satisfies the following: $f(\mathcal{B}^{\text{e}})=\gamma,$ where $\gamma$ ---
continious curve: ${R}\rightarrow\mathcal{N},$ such as $\pi_1\circ\gamma=\mathcal{T}$
and
$\pi_1|_\gamma$ --- homeomorphism:
\begin{equation}\label{def1}
\pi_1|_\gamma\in\text{Hom}(\gamma,\mathcal{T}).
\end{equation}
will be referred to as {\it a family of elementary hyperbodies.}
In other words, elementary hyperbodies are inverse images of ordinary mass points'
world lines.

\medskip
{\bf The Proposition 4.1}  \tt
Elementary hyperbodies define $\mathfrak{P}$-structure in regard to the
 group of transformations
$\mathcal{G}_{\mathcal{T}}.$

\smallskip
\noindent{\bf Proof.} \rm Let $\mathcal{G}^{e}$ --- group of invariance of
elementarity properties.
Let $\gamma$ --- observable image of some elementary hyperbody in
$\mathcal{N}$ and $\gamma_g\equiv g(\gamma)$ --- its image under
action of $g\in\mathcal{G}^e.$
By definition of  $\mathcal{G}^e$ and elementarity property we have the following
chain of equalities:
\begin{equation}\label{ident}
\mathcal{T}=\pi_1\circ\gamma_g=\pi_1\circ g\circ\pi|_\gamma^{-1}\,\mathcal{T},
\end{equation}
that gives $\pi_1\circ g\circ\pi|_\gamma^{-1}\in\text{Aut}(\mathcal{T}).$
It means, that $g$ transform fibers of simultaneous events to itself, i.e.
$\mathcal{G}^e\subseteq\mathcal{G}_{\mathcal{T}}.$ Inverse inclusion
follows from (\ref{ident}) together with continuity of projections
and Lie group action.$\Box$

\smallskip\noindent
In our case (\ref{def1}) ---
is a determining property of elementary hyperbodies and  $\mathfrak{A}=
\mathcal{T}, D=1.$
Elementary hyperbodies by definition allow us to consider {\it physical  hyperbodies}
$\{\cup_{\alpha}\mathcal{B}^{\text{e}}_{\alpha}\}.$

%2) Bodies' histories' continuity implies continuity of their evolution in $\mathcal{V}$-space.
%Formally, for any elementary hyperbodies' pair $\mathcal{B}^{\rm
%e}_{1,2}\in\mathcal{B}$   it occurs:
%\begin{equation}\label{cont}
%\eta(W_1,W_2)\in C^0(\mathcal{T}),
%\end{equation}
%where $W_{1,2}=f_{\mathcal{V}}(\mathcal{B}^{\rm e}_{1,2}).$
%Due to the continuity of metric,
%its projection and composition with any diffeomorphisms, we are making a conclusion
%that the following holds true:
%
%{\bf Statement 6.2}
%{\sl The property of histories' continuity defines $\mathfrak{P}$-structure with regard to
%common group of observers' diffeomorphisms  ${{ \mathcal{G}}}$.}

2) The possibility of determining physical time is connected with the possibility of
measuring it using special bodies with hyperbodies, whose observable
evo\-lu\-ti\-on we call {\it periodic.}  Let $\text{Per}(\mathcal{T})$ --- a family of
con\-ti\-nu\-ous periodic functions, determined on $\mathcal{T}.$
Let us define a family of  "standard"\, hyperbodies
$\{\mathcal{B}^{\text{per}}\},$ such as for any $\mathcal{B}^{\text{per}}$
and any $\mathcal{B}^{\text{e}}_1,\mathcal{B}^{\text{e}}_2
\in\mathcal{B}^{\text{per}},$ it occurs:
\begin{equation}\label{periodic}
 \eta(b_1^T,b_2^T)\subset{\rm Per(\mathcal{T})},
\end{equation}
where $b_{1,2}^T=\pi_2(\mathcal{V}_T\cap f(\mathcal{B}^{\text{e}}_{1,2}))$ for all
$T\in\mathcal{T}.$

\medskip
{\bf The Proposition 4.2}\  \tt The family of standard hyperbodies $\{\mathcal{B}^{\rm per
}\}$ defines $\mathfrak{P}$-structure with re\-gard to the group ${A(1,{R})}\times {{\mathcal{G}}}_{\eta}\subset
\mathcal{G}_{\mathcal{TV}},$ where ${A(1,{R})}$ -
nonhomogeneous affine group acting on co\-or\-di\-na\-te space $\tau:\,\mathcal{T}\to R,$ and ${{
\mathcal{G}}}_{\eta}\le {{ \mathcal{G}}}$ - isometry group of metric
$\eta.$

{\bf Proof.} \rm Let for some physical body (\ref{periodic}) takes place.
It means, that lefthand side of  (\ref{periodic}) can be represented as the following
Fourier row:
\[
\eta(b_1^T,b_2^T)=\sum\limits_{n=0}^\infty a_n e^{in\omega\, \tau(T)},
\]
where $\tau(T)$ --- perceptive time coordinate, $\omega$ ---
some main frequency. We can rewrite this relation in terms of
transformed by $g=(g_0,g_\eta)\in A(1,R)\times {{\mathcal{G}}}_{\eta}$
values: $T'=g_0(T),$ ${b'}^T=g_\eta(b^T)$:
\[
\eta(g_\eta^{-1}({b_1'})^{g_0^{-1}(T')},g_\eta^{-1}({b_2'})^{g_0^{-1}(T')})=
\sum\limits_{n=0}^\infty a_n e^{in\omega\,  \tau(g_0^{-1}(T'))}.
\]
By definition of $g_\eta$ together with  rule $\tau\circ
g_0(\bullet)=\alpha\tau(\bullet)+\beta,$ ($\alpha,\beta$ --- affine group parameters)
we obtain:
\[
\eta(\tilde b_1^{T'},\tilde b_2^{T'})=\sum\limits_{n=0}^\infty a_n'e^{in\omega' \tau'(T')}
\]
--- relation, similar to  (\ref{periodic}). Here
$\tilde b_i^{T'}=\pi_2(\mathcal{V}_{g_0^{-1}(T')}\cap f_g(\mathcal{B}^{\text{e}}_i)),$
$a_n'=a_ne^{-in\omega\beta/\alpha},$ $\omega'=\omega/\alpha.\,\Box$

3) We are fixing natural partial order "$\prec$"\, on manifold $\mathcal{N}$
in the following way:
\begin{equation}\label{order}
P_1\prec P_2\ \Leftrightarrow  \tau(P_1)\prec \tau(P_2)\quad\text{for
all}\quad
P_1,P_2\in\mathcal{N},
\end{equation}
where $\tau(P_1), \tau(P_2)$ --- absolute (perceptive) time
co\-or\-di\-na\-tes of events $P_1,
P_2$, ordered as elements of  $R.$
%where $X^0(P_1), X^0(P_2)$  - absolute temporary coordinates of the events $P_1,
%P_2,$ which are ordered as the points of real line ${R}.$

{\bf The proposition 4.3} \tt Partial order "$\prec$"
defines $\mathfrak{P}$-\\ structure with regard to
subgroup $\mathcal{G}_{\mathcal{T}}^+\subset \mathcal{G}_{\mathcal{T}},$ for which
$d\tau_\mathcal{G}/d\tau>0.$

{\bf Proof.} \rm Let $\tau_g=\tau\circ\pi_1\circ g$ --- result
of action of order group symmetry, representing in local coordinates.
Using identity:
\[
f(x)-f(y)=\langle\overline{\partial f},(x-y)\rangle,
\]
where $x,$ $y$ --- points of  $R^n,$ $f$ --- smooth function $R^n\to R,$
$\overline{\partial f}$ --- overaged over segment $xy$ covector
$\partial f/\partial\xi_i,$ $\langle\, ,\, \rangle$ --- standard
pairing of forms and vectors,
we get:
\begin{equation}\label{overg}
\tau_g(P_1)-\tau_g(P_2)=\alpha_0(\tau_1-\tau_2)+\alpha_i(X_1^i-X_2^i)>0,\quad
\end{equation}
$(i=1,2,3).$ Here $\{X^i_{1,2}\}$ --- local  coordinates
of $P_1$ and  $P_2$ on $\mathcal{V},$
$\alpha_\mu=\overline{\partial\tau_g/\partial X^\mu},$ $(\mu=0,1,2,3),$ $X^0=\tau.$
If  $\alpha_i\neq0,$ then by suitable choice of space perceptive
coordinates (\ref{overg}) can be always violated. So, requirement  of ordering
conservation leads to the condition $\alpha_i=0.$ By continuity of partial derivatives
it gives $\partial \tau_g/\partial X^i=0.$
It means, that we are restricted by  $\mathcal{G}_\mathcal{T}.$
For conservation of  (\ref{overg}) it is also necessary, that for all   $P_1$
and  $P_2$ the inequality $\alpha_0>0$ be satisfied. By  continuity of partial derivatives
it gives: $d\tau_g/d\tau>0.\,\Box$

Note, that as against the previous $\mathfrak{P}$-substructures,
we are introducing time arrow here irrelatively of  pro\-per\-ti\-es of any
"asymmetrical in time"\, physical bodies and therefore time arrow
is an inner property of an
observer\footnote{Of course, we can say that the observers themselves are
a special family of "bodies", each member of which is capable of inducing time arrow (see also \cite{pav}).}.
This is mainly connected with well known fact that
basic laws of classical mechanics are invariant with regard to time inversion.

Thus, complete $\mathfrak{P}^{\text{chron}}$-structure of time is given by family
of elementary hyperbodies and
their unions with properties of continuity, periodicity, and time order.
Appropriate common subgroup $\mathcal{G}^{\text{chron}}\le \mathcal{G}$
is an in\-ter\-sec\-ti\-on of all subgroups, defined by $\mathfrak{P}$-substructures it is made
up of and has a form of direct product
${A}^+(1,{R})\times \mathcal{G}_{\eta}.$

\subsection{$\mathfrak{P}$-geometry of physical time}

Direct product, obtained in previous paragraph,
belongs to $\mathcal{TV}$-class of diffeomorphisms
and each multiplier is in itself a normal divisor of  direct product.
$\mathfrak{P}^{\text{chron}}$-geometry,
corresponding to the ${\mathcal G}_{\eta},$
relates to 3-di\-men\-si\-o\-nal space of classical physics and turns out to be
"very indistinct"\, at this stage.
%$\mathfrak{P}^{\text{chron}}$-geometry of space will be considered in
%more detail in the second part of the article.
Experimental approach to a visual 3-geometry  investigation we'll consider
in the second part of the paper.
For the moment being we are going to concentrate on analyzing geometry of physical time,
contained in affine group ${A}^+(1,R).$
Its group space is an open two dimensional manifold, homeomorphic
$R^{+}\setminus\{0\}\times R,$
with cut edge. For any two elements ${a},{b}\in{A^+(1,R)}$
group composition law is: ${a}\circ{b}=(a_1,a_2)\circ(b_1,b_2)=(a_1b_1,a_1b_2+a_2).$
Let us determine generating function $\rho_0$ for group $A^+(1,R).$
Isometry condition (\ref{isometry}) becomes:
\begin{equation}\label{isA}
\rho_0(\xi x_1,\xi x_2+\eta,\xi y_1,\xi y_2+\eta)
=\rho_0(x_1,x_2,y_1,y_2),
\end{equation}
where $(\xi,\eta),\, (x_1,x_2),\, (y_1,y_2)\in{A}^+(1,R).$
Differentiating both parts of (\ref{isA})  by $\xi$ and $\eta,$
we are getting defining system of differential equations of the first order:
\[
x_1\frac{\normalsize\partial\rho_0}{\normalsize\partial
x_1}+x_2\frac{\partial\rho_0}{\partial x_2}+y_1\frac{\partial\rho_0}{\partial
y_1}+y_2\frac{\partial\rho_0}{\partial y_2}=0;\quad
\frac{\partial\rho_0}{\partial x_2}+\frac{\partial\rho_0}{\partial y_2}=0.
\]
Their common solutions respectively are:
\[\rho_0=f\left(
\frac{x_2}{x_1},\frac{y_1}{x_1},\frac{y_2}{x_1}\right);\quad
\tilde\rho_0=h(x_1,y_1,x_2-y_2),\]
where $f$ and $h$ --- arbitrary differentiated functions. Uniting them, we finally get:
\begin{equation}\label{genA}
\rho_0=\rho_0\left(\frac{y_1}{x_1},\frac{x_2-y_2}{x_1}\right).
\end{equation}
Consider two independent generating functions:
\[ \rho_{01}=\frac{y_1}{x_1};\quad\rho_{02}=\frac{x_2-y_2}{x_1}\]
and check them for metricity.

{\bf The Proposition 4.4} \tt Generating function $\rho_{01}$ satisfies metricity condition (\ref{crit})
only on sec\-ti\-ons ${const}\times R$ of group space
${A}^+(1,R)$, where
$\rho_{01}\equiv1.$

{\bf The Proposition 4.5} \tt Generating function $\rho_{02}$ satisfies metricity condition (\ref{crit})
in two and only two cases:
\begin{enumerate}
\item on sections ${const}\times R,$  with
$\rho_{02}={const}\cdot(y_2-x_2).$
\item
on sections $R^+\times{const},$  with $\rho_{02}\equiv0.$
\end{enumerate}
\rm Excluding trivial cases $\rho={const},$ we are getting,
basing on propositions of section \ref{teor},
to the following general expression of physical $\mathfrak{P}^{\text{chron}}-$metrics:
\begin{equation}\label{Ametr}
\rho(t_1,t_2)=\varphi(|t_2-t_1|),
\end{equation}
where $\varphi$ --- differentiable function, satisfying condition of proposition 3.4.
Most important in practical sense and having technically the simplest construction, metrics class
from (\ref{Ametr}) is given by:
\begin{equation}\label{Aexample}
\rho(t_2,t_1)=C|t_2-t_1|^{\alpha},\, 0<\alpha\le1.
\end{equation}
From the physical viewpoint constant $C$ defines physical units of
time. Linear Euclidean case, commonly used in classical mechanics is obtained from
(\ref{Aexample}) under $\alpha=1$.

\section{Experimental research of visual manifold geometry}
\subsection{Experiment description}

The idea of determining quantitative characteristics of visual manifold
geometry\footnote{Instead of symbolic term "space"\, we are going to use more exact
term "manifold"\,
in the second part.} is simple:
it is necessary to find correlation between geometrical
characteristics' estimates by statistically considerable number of ob\-ser\-vers
for given objects and true characteristics of these objects, determined by measuring instruments.
Despite its sim\-pli\-ci\-ty, practical realization of this idea is quite
a labor consuming problem. It is worth noting here just a few general problems arising on
our way along with some significant details of the carried out experiment.
\begin{enumerate}
\item
In visual manifold geometry, there is a natural length scale $\ell$ ---
observer's own size. Because of this, whole visual manifold is naturally divided
into charts $\{\mathcal{V}^{\rm near}_i\}$, $\{\mathcal{V}^{\rm mid}_j\}$, $\{\mathcal{V}^{\rm far}_k\},$
covering  {\it near} ($d\ll\ell$) {\it middle} ($d\gtrsim\ell$) and  {\it
far} ($d\gg\ell$) zones correspondingly, where $d$ ---
characteristic distance from observers' eyes to some
internal point within a map\footnote{We are assuming here that characteristic map
size is much less than a distance between a map and observers'
eyes.}. In our experiment we, due to technical reasons, have
limited ourselves to middle zone: $d\sim 1.5$m--$15$m.
\item
Complete research of visual manifold's geometrical properties comprises of
studying its at least three relatively independent aspects: topology,
con\-nec\-ti\-on  and metrics.
In the present work we have only studied metric  properties,
since they do not require sophisticated equipment for their me\-a\-su\-re\-ment and are more common.

\item
Since visual manifold is a part of observer's united perceptive manifold,
it is natural to expect that geometrical characteristics of an ob\-ser\-ved object will
be influenced by the properties of those ob\-ser\-ved objects that pertain to other
subspaces of per\-cep\-ti\-ve space (for example, object's color which is object's
non-geometrical characteristic). To exclude those "undesirable interactions"\, of
different perceptive subspaces from our work, we have prepared the artificial objects
of unified format --- 8 blue rec\-tang\-les on white background differing in
sizes
and proportions. They have been shown to the probationers -- "observers"\, -- under relatively
the same conditions (university's auditorium) at various distances. The exact dimensions
of the rectangles are given in Table 1 of Appendix.
The observers have been proposed to estimate the samples dimensions and the distance to them.
After that, they have been asked to enter their estimates into special card.
%\item
%To get quantitative metric characteristics for visual manifold, it is necessary to use a
%certain "background"\, (reference) geometry, "an observer"\, should lean on, while estimating
% objects' dimensions. It is natural for "the observers"\, to use as such "solids geometry"\,
%  --- Euclidean geometry though, according to Poincare's ideas, other
% variants are theoretically possible.
\item To eliminate subjectivity at estimating objects' linear sizes, it's been necessary
to question quite a number of the "observers"\, to smoothen individual perception
peculiarities after averaging data and, on the contrary, to make common characteristics of
 visual manifold geometry appear. Due to tech\-ni\-cal reasons, only 80 observers
  have been involved in our experiment. As our study has shown, this value is,
in fact, a lower limit at which general laws of  perception start to show.
 When processing the results, we have ignored  the data, obtained from the same
individual, but severely fluctuated around object's true dimensions and distances to them.
We have interpreted the situations like this as influence by purely psychological factor ---
such "observers"\, have not been very responsible to our experiment\footnote{The
majority of "observers"\, have been graduate students of YSPU who get a little tired at
the end of the day. Fortunately, we have faced the situation described above only a
 few times.}.
\item
It is well known to neurophysiologists  that  different areas of brain crust are
 responsible for  perception of  vertical and horizontal dimensions \cite{hubel}. This tips to the idea
  that metric properties of visual manifold can discover anisotropy. To be able to study
it (anisotropy), we have distinguished (for ourselves) between objects' orientation and
their vertical and horizontal dimensions.
\end{enumerate}

In the course of experiment, the students have been divided into groups each
consisting of 16 people. Each group's members have occupied one row consisting of 8 tables each.
The tables have been at fixed distances from the showed rectangles. These distances can be
found in Table 2 of the Appendix. All students have been given the cards.
 The cards had objects' numbers on them and opposite each object's number there have been
three blank spaces designated by letters V, H, and D standing for "estimated vertical
dimension", "estimated horizontal dimension", "estimated distance to the object"\,
correspondingly. Having been shown an object and having fixed its' dimensions in the card,
the students circularly changed their seats according to the following scheme:
$1\to2\to3\to4\to5\to6\to7\to8\to1.$ Thus, each group has finally given two estimates on each object's
characteristic and at each distance to the object. We have overall tested about five
 such groups so on each object's characteristic and at each distance to the object we
  got sampling consisting of 10 estimates we could later do averaging on.
  In the Appendix the  Table 3 of the averaged experimental data
  is presented.

Basing on experimental data summarized in the table,
we have drawn the dependencies $\overrightarrow{X}(\overrightarrow{x}),$
where $\overrightarrow{X}=\{H,V,R\}$
and $\overrightarrow{x}=\{h,v,r\}$ --- vectors of estimated and true characteristics ---
heights, widths, and distances to the objects correspondingly.
The interpretation of the obtained dependencies has been done on the basis of  {\it affine
model} of visual manifold geometry. It is assumed within the frames of this model that
the averaged data, obtained from the students, are linear functions on objects'
 true characteristics:
\[X_i=k_ir+X_{i0},\quad i=1,2,3,\]
where $k_i$ -- {\it linear radial overstating
coefficient of width $(i=1),$
height $(i=2)$ and of a distance to the object $(i=3).$}
Parameter $X_{i0}$ --- ideal parameter, having a sense of correspondent dimension estimate
"at zero distance". In some sense it is more convenient to analyze a behavior of differences
$\Delta_i=X_{0i}-x_i$ and dimensionless ratios $\varepsilon_i=\Delta_i/X_{0i}.$
For the most dependencies it is possible to introduce one important
characteristic $r^\ast$
---
{\it ideal distance,} defined from the equation: \[
\overrightarrow{X}(r^\ast)=\overrightarrow{X}_{\text{true}},
\]
i.e. such a distance
at which objects' estimated cha\-rac\-te\-ris\-tics coincide with the true ones.
Finally, let us introduce one more notion --- {\it radial  anisotropy coefficient:}
$\sigma=1-k_{2}/k_1,$ where $k_1$ --- radial overstatement coefficient for objects'
horizontal characteristics, and $k_2$  --- for the vertical ones.
% Consider the most interesting dependencies $\vec X(r)$ of
% estimated linear dimensions' from distances. It is easy to see that on nearly all
%  of them (except for $R(r)$) there is an increase from understated estimate on the left
%(lower end of distance range), to overstated one at longer distances (Tables 3,4).
%In other words, in linear relation

All of the obtained characteristics for linear model are shown in overall Table 4 in the
Appendix. Let us analyze the acquired results setting up the following rather natural
 assumptions from our daily experience as trial hypothesis:
\begin{enumerate}
\it
\item
For objects having  sides proportion near to 1 and large area, their dimensions are overstated,
and the distances to them are understated;\label{1}
\item
The dimensions of little objects at long distances are understated, and the distances from
them --- are overstated;\label{2}
\item
Perception of one side for substantially ani\-so\-tro\-pic objects influences the perception of
the other;\label{aniz}
\item
It is possible that there exist some stable proportions, perception of which
is sufficiently different from that of the other proportions.\label{proper}
\end{enumerate}
Basing on table 4 data, it can be checked whether the above statements are right.
First two lines of the table clearly show that despite hypothesis \ref{1} and \ref{2}
radial sizes overestimation coefficient is positive for all objects except the object 6.
This coefficient is particularly great for objects 3,4,7.
The difference between vertical and horizontal overestimation coefficients for anisotropic
objects 3,  6 and 7 verifies hypothesis \ref{aniz} and corresponds to its
 particular case when greater side is being more times overestimated.
For the symmetric object 4 overestimations coefficients practically coincide
(it supports hypothesis \ref{1} and \ref{aniz}).

 Practically exact
 coincidence of overestimation co\-ef\-fi\-ci\-ents for symmetrical (or nearly symmetrical) objects
  4 and 8 is natural. Objects 1 and 2 also reflect aforementioned variant of
 hypothesis \ref{aniz}, but to weaker degree. Objects 5 and 6 are exclusions.
 First of them has a ratio between overestimation coefficients which corresponds to
 inverse variant of hypothesis 3: greater side is being overestimated by fewer times.
 The second one has been underestimated in both vertical and horizontal sizes, and, what
 is interesting,  this is according to direct variant of hypothesis \ref{aniz}.
It's worth noting that both objects have largest areas
(approximately 103 $\text{cm}^2$ and 140 $\text{cm}^2$ correspondingly) and proportions of the first are
close to 3:2 (exactly 1.57), and the second's --- to golden section $(3-\sqrt{5})/2\approx0.381$
(exactly 0.369). We are going to make sure further that objects 5 and 6
possess a number of other "strange"\, pro\-per\-ti\-es, apparently according to hypothesis \ref{1} and
\ref{proper}.

The table's bottom line reflects the above mentioned as well. Besides, in this line there
have been unexpectedly strong anisotropy for perception of  vertical object 7
having sides proportions close to 1:2.

As against radial over\-es\-ti\-ma\-ti\-on coefficients of objects' sizes, pertaining to their
picture plane, radial over\-es\-ti\-ma\-ti\-on coefficients of distances $k_3-1$
are all negative but the smallest object 8.
In full accordance with hypothesis 1, the objects 3,5,6
with largest areas, have maximum underestimation coefficients.
Objects 1,7,8 with smallest areas, have minimum un\-der\-es\-ti\-ma\-ti\-on coefficients,
where object 8 verifies hypothesis \ref{2}. %Object 2, despite its small area,
%has an underestimation coefficient equal to that of the object 6
%with medium area. Here, again we are apparently facing shape (form) anisotropy effect:
%tall and narrow objects appear closer than they are.

Estimates at zero distance $\Delta_i$ all turn out to be either negative or null.
In conjunction with the first two lines, this implies that nearly all objects have
positive "ideal distance". Various dimensionless estimates $\varepsilon_i,$
correspond to anisotropic objects, and isotropic objects are being corresponded by "close"\,
ones.  Objects 5 and 6 are again the exclusions. The first one has $\varepsilon_1>0,
\varepsilon_2<0,$  and the second --- $\varepsilon_1=0,\varepsilon_2=0.12.$
This anomaly, along with the first two, leads to anomalously  big negative "ideal distances"\,
for object 5 and to anomalously big positive ones for object 6.

Parameter $R_0$  fluctuates around zero and is in\-sig\-ni\-fi\-cant for all objects.

Symmetrical (4,8) or almost symmetrical (1,7) objects
obviously have an "ideal distance"\, (4-5 m), at which the perception of their vertical
and horizontal dimensions approximately corresponds to their real values.
It is worth mentioning that inaccuracy in radial perception remains.
In accordance with hypothesis \ref{aniz} the most significant
divergence in ideal distances takes place for anisotropic objects: 2 and 3
and anomalous object 6. Anomalous object  5 "has no"\, horizontal ideal distance at all.
As to ideal distance to estimate the distances themselves, it does exist
(i.e. notable exceeds zero) only for the smallest object 8 and equals, surprisingly, to 28m.

Thus, we can conclude for all radial dependencies  with the following remarks which
 substantially specify and complete our initial a priori hypothesises:

\begin{enumerate}
\it
\item
Estimates for distances to the objects mostly satisfy hypothesis  \ref{1}
and \ref{2}.
\item
Estimates for horizontal and vertical dimensions do not generally satisfy
hypothesis \ref{1} and \ref{2} in a sense that almost all radial overestimation
co\-ef\-fi\-ci\-ents exceed zero and satisfy hypothesis in other sense: big
dimensions are overestimated to a bigger extend and the smaller ones ---
to a smaller, according as well to general hypothesis \ref{aniz}.
\item
Objects 5 and 6, having proportions 3:2 and "golden section"\,
(and probably 7 with proportion 1:2),
and largest areas, fall out of general law by a number of factors and evidently play
a special role in geometry of perception in accordance with hypothesis \ref{proper}.
\item
There is an ideal distance (4-6m) for symmetrical or almost symmetrical objects.
At this distance, their perception adequately reflects their real di\-men\-si\-ons.
\item
The perception of extended objects is distorted at any distances.
\end{enumerate}

Let us briefly describe the characteristics of visual manifold basing on other relations
 obtained from Table 4. The relations between perceived proportions of sides and their true
 proportions are for all distances well approximated by linear function with a
 coefficient being a little less or equal to 1. Since, proportion is a ratio of horizontal
 dimension to vertical one, the result we have obtained points to the fact that in general,
 vertical and horizontal perception "distortions"\, are a little bit different.
 Horizontal dimensions are on the average overestimated a little less than the vertical.

3D graphs of perceptive width and height de\-pen\-den\-ci\-es on their true
values are presented in Figure 1 (see Appendix).
Comparison of this two graphs support conclusion about
anisotropy of visual manifold. In first, distortion of
vertical sizes perception are expressed more strongly, than distortion of
horizontal sizes perception. This is illustrated by horizontal
sections of these surfaces in Figure 2.
In second, shape of distortions of vertical and horizontal perceptive
geometries from Euclidean geometry are different. We illustrate it in  Figure 3.
It presents\footnote{Dependencies, represented in Figure 3 are nonlinear,
so in what follows we discuss some general features of \it nonlinear perceptive
space model.} pairs of 2-dimensional dependencies:
ex\-pe\-ri\-men\-tal (bright) taken from Figure 1,
and Euclidean (dark) of type $H=h$ и $V=v$.
Boundaries of bright and dark areas  ---  "ideal"\, curves, i.e. going through
a set of parameters, for which true and perceptive objects' characteristics coincide.
We see, that
that while horizontal characteristics are described by
a weakly up-convex "perception surface"\,,
vertical characteristics are des\-cri\-bed by
apparently down-convex "perception surface".
Thus, we conclude, that vertical estimates have no "ideal"\, range for perception (height)
(are always overestimated), while horizontal ones
this "ideal"\, range is about 3-5m
and horizontal sizes about  5-15cm.

We would like to stress that all our conclusions are approximate and require further
correction (for example, by increasing testers' number).%   3-dimentional charts featuring width and height relative perception mistake relating
% to width or height and distance are shown in pic.\ref{r3d}.
%Comparing these two charts once again proves a conclusion regarding visual manifold's
%anisotropy. Particularly, it is easy to see that on the first chart for widths,
%sections $h={\rm const}$ have a maximum which is moving towards decrease in $r$
%starting from 6m for objects with the largest width. Something similar is observed
%for vertical mistake, besides, there is a clearer expressed maximum for sections $r={\rm const},$
%which is also drifting towards increase in heights as $r$ grows.
%Thus, the most "uncomfortable"\,
%element for perception can be emphasized for heights: maximum relative
%mistake in perception (about 0,2) corresponds to distances 5-6 m and heights 18-20 cm.

\section{Conclusion}

We have made an attempt basing on Poincare's ideas to find a tie
connecting perceptive space geometry,
physical objects, and physical geometry which is used when
formulating and analyzing the laws of  physics.
Basic ideas of this approach have been demonstrated using physical
chronogeometry as an example. Experimental research of visual perceptive
space geometry's pe\-cu\-li\-a\-ri\-ti\-es has shown its' non trivial character even
within the limited area (middle zone)  that attracted our attention.
Obviously, our affine model should be considered as linear approximation of more
general, nonlinear model  which is necessary for obtaining complete geometrical
picture of the visual perceptive space.

Let us make some general remarks in conclusion.

\begin{enumerate}
\item
As it can be seen from the basic points of the approach  being set forth,
observer's conception is necessary for building physical picture of the world
even at the level of classical physics. Without an observer --- his
Newton's mapping, world and perceptive manifolds remain absolutely detached and isolated.
In other words, an observer is built in the surrounding world in such a way that he
is not only (and not so much) a passive spectator but an active participant in forming and
uncovering laws of physics.
\item
Despite this circumstance, and in spite of the thing that the geometry of
perceptive space should apparently play its role at any stage of our physical
reasoning, the laws of the nature can be formulated in such a manner as if
there was no ubiquitous  geometry-mediator at all. For instance, when de\-du\-cing
chronogeometry metrics (\ref{Aexample}), at the interim stage of the argumentation
we only used a fact of existence of some perceptive geometry,
but its specific pro\-per\-ti\-es turned out to be unsubstantial for deducing formula (\ref{Aexample}).
Evidently, this situation is typical for all of the laws of classical physics of XIX century,
in which the role of the observer is disguised by the notions of absolute space,
time and a number of others. Some sort of the physical laws' dependence on an
observer starts to reveal itself in SR and GR.
The role of the observer in these theories is played by the the notion of the reference
frame\cite{mitsk}\footnote{SR  and GR conception of the reference frame is a
particular case of Newton's mapping formalism. In SR and GR we have
 tensor  bundle  of Riemannian  manifold as world manifold, in a capacity of an observer ---
 bundle of orthonormal tetrads over this manifold, in a capacity of Newton's mapping ---
 projection of  physical or geometrical tensors (objects of tensor bundle  over the said manifold)
 onto tetrads vectors and 1-forms dual to them.}.
 Likewise, within Maxwell's electrodynamics in Minkowski space-time, angular velocity
 of  the reference frame's rotation can imitate magnetic charges' density;
 in Friedmann-Robertson-Walker's cosmology,  the observed volume of the universe can be
 both finite and infinite depending on reference frame even within the frames of the
 same cosmological model. Though, covariant approach, reflecting, apart from invariance idea,
 some general scientific-philosophical  aims of the modern scientific thinking,
 again takes observer's perceptive space out of context  when formulating the laws of
 relativity physics. The further development of our approach into the field of
 quantum phenomena will probably enable to attribute some part of "strangenesses" and
 oddities  of microworld to the observer's perception geometry (see, for example, \cite{qrf}).
 The detailed evolution and development of our approach into relativity physics area
 could as well serve as a foundation for the new interpretation of some observation
 cosmology facts.
\item
In the present article we have just touched upon space-time aspect of perceptive space.
The said space, taking into account the full range of sen\-sa\-ti\-ons and perceptions,
is much broader and includes auditory, haptic, motor, gustatory, olfactory,
ther\-mal and a number of the other subspaces which, in addition,
are connected with each other by the complex, as a matter of fact,
non-functional de\-pen\-den\-ci\-es. We are convinced that the combined research of the
perception space by the methods of mathematics, physics and traditional sciences
about a human being could be a durable basis for building up a uniform language for
describing different phenomena of the outer (surrounding) world and a man ---
as its in many respects unique representative.
\end{enumerate}

%\section{Figures}
%
%Please include the figures into your paper as follows
%
%%%%          EITHER
%
%\Figure{fig1.eps}   %% for one-column figures
%       {Caption of the figure}
%
%%%%          OR
%
%\WFigure{fig1.eps}   %% for two-column figures
%        {Caption of the figure}

%\section{On the references}
%
%References are numbered in the text in the form [1], [2], etc.;
%the format of the list of references is indicated in the sample that
%follows, so please carefully follow it, paying attention to the order
%in which the information appears, spaces, quotation marks, etc. The
%titles of articles are (unlike books) not necessary, but desirable for
%preprints and e-prints.
%
%The order of references in the list may either correspond to first
%citations in the text, or be alphabetic.
%
%\Acknow
%{The authors are grateful to the Editorial Board for accepting their
%graphomania for publication.}

%\begin{thebibliography}{99}
%
%\bibitem{1}
%     J.J. Jones, {\it Phys. Rev.\/} {\bf U 999}, 9999 (1999).
%\bibitem{2}
%     G. Green, B. Blue and W. White, ``Theory of Everything'',
%     North Pole Univ. Press, North Pole, 1900.
%\bibitem{3}
%     A.A. Ivanov and B.B. Petrov,
%     ``The Universe and Its Neighbourhood''.
%     gr-qc/9909999; to appear in {\it Phys. Rev. Z}.
%\bibitem{4}
%    B.B. Petrov, {\it in:\/} ``Selected Discoveries'', ed. A.A. Ivanov
%    et al., North Pole Univ. Press, North Pole, 1999.
%\bibitem{5}
%    F.F. Sidorov, ``Tops of Gravitational Science'', PhD
%    thesis, Everest University, 1999.
%\end{thebibliography}

\onecolumn
\appendix
\section{Figures}

%\vspace{-8em}

%\refstepcounter{figure}\label{r3D}
\begin{tabular}{ll}
\includegraphics[width=.37\textwidth, height=0.37\textwidth]{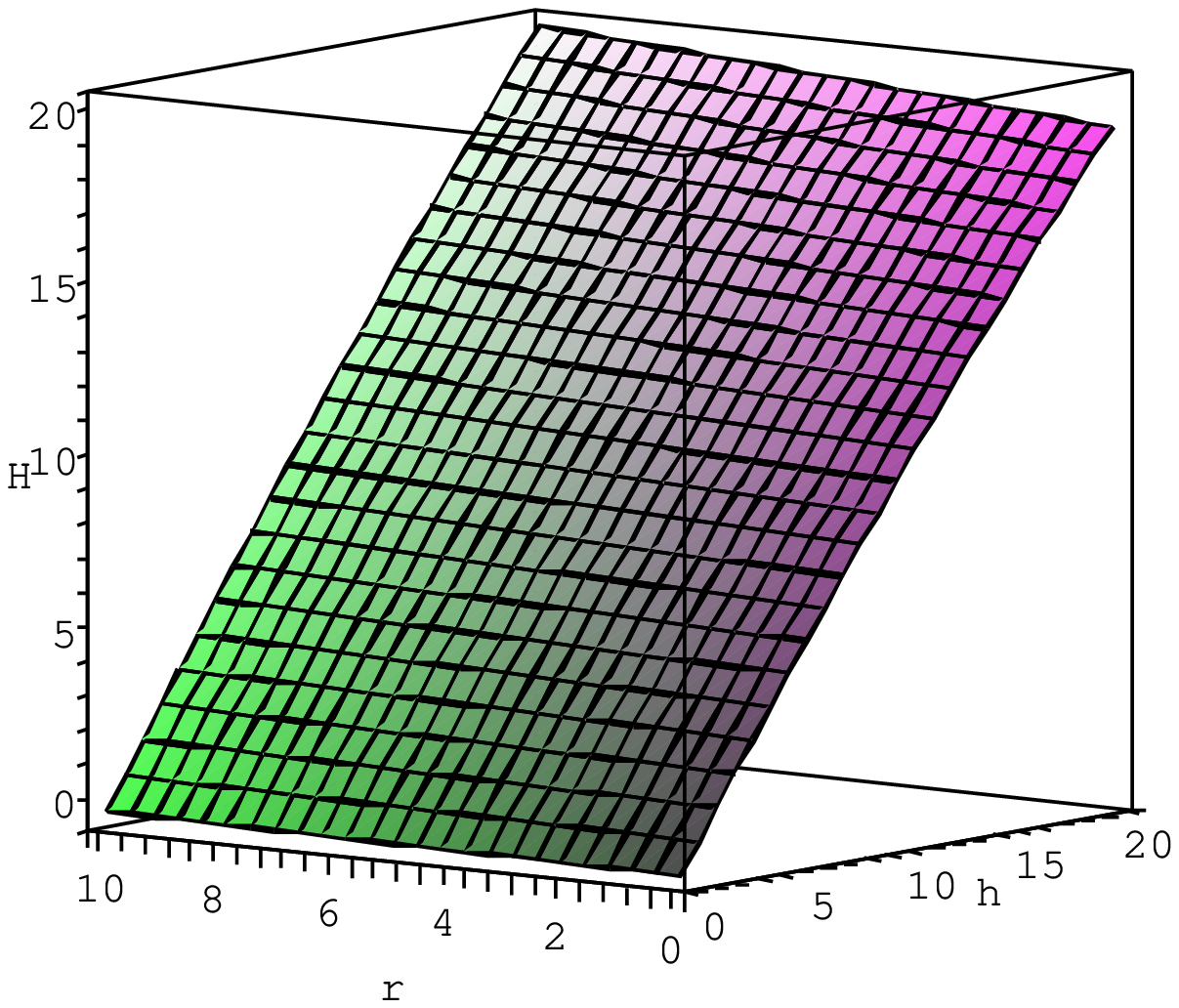}&
\includegraphics[width=.37\textwidth, height=0.37\textwidth]{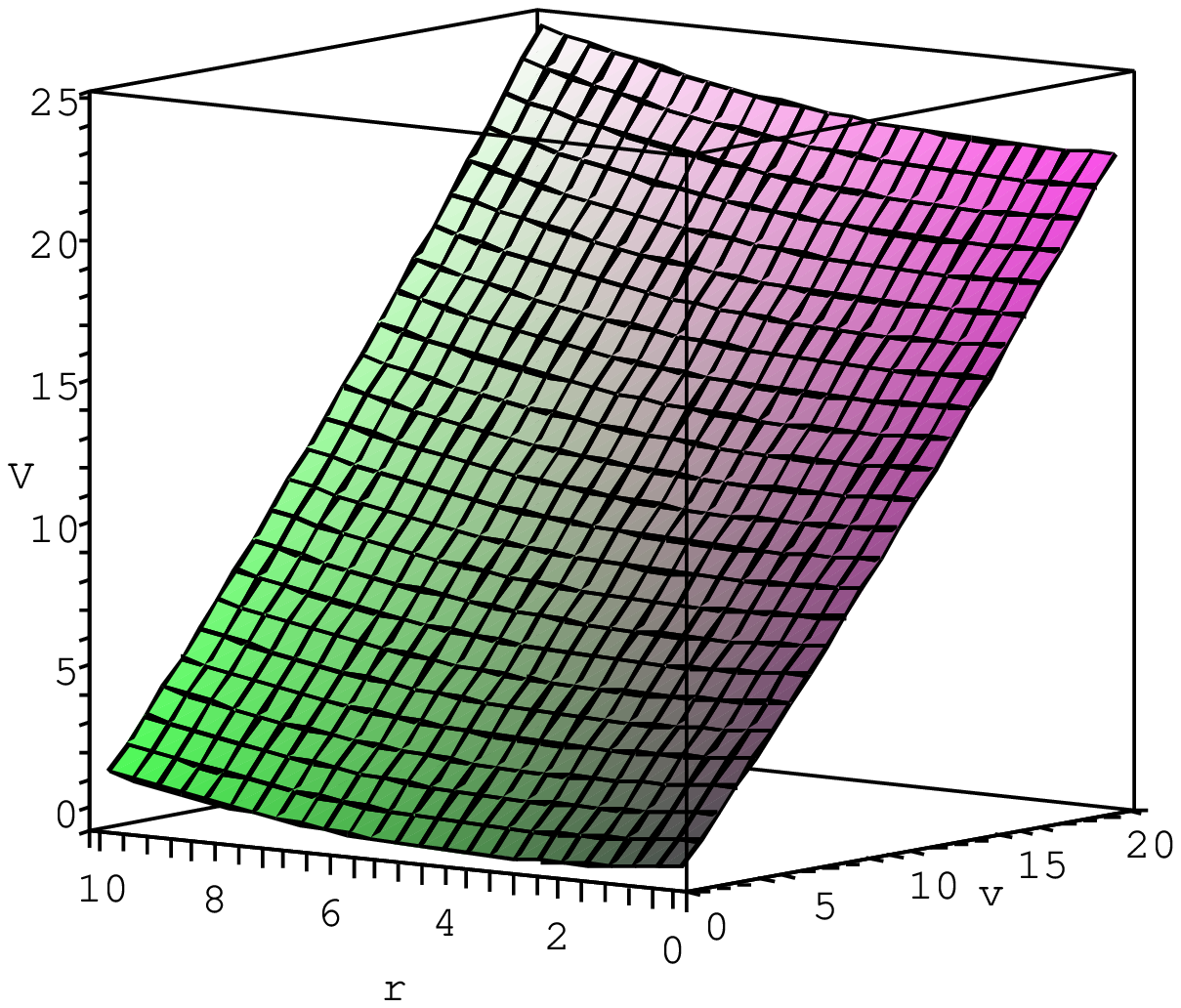}
\end{tabular}

{\small Figure 1. 2-dimensional dependencies of perceptive parameters (width and height)
on correspondent  true parameters and distant to the objects.
%\hfill
}

%\vspace{-3em}

%\refstepcounter{figure}\label{r3d1}
\begin{tabular}{ll}
\includegraphics[width=.37\textwidth, height=0.37\textwidth]{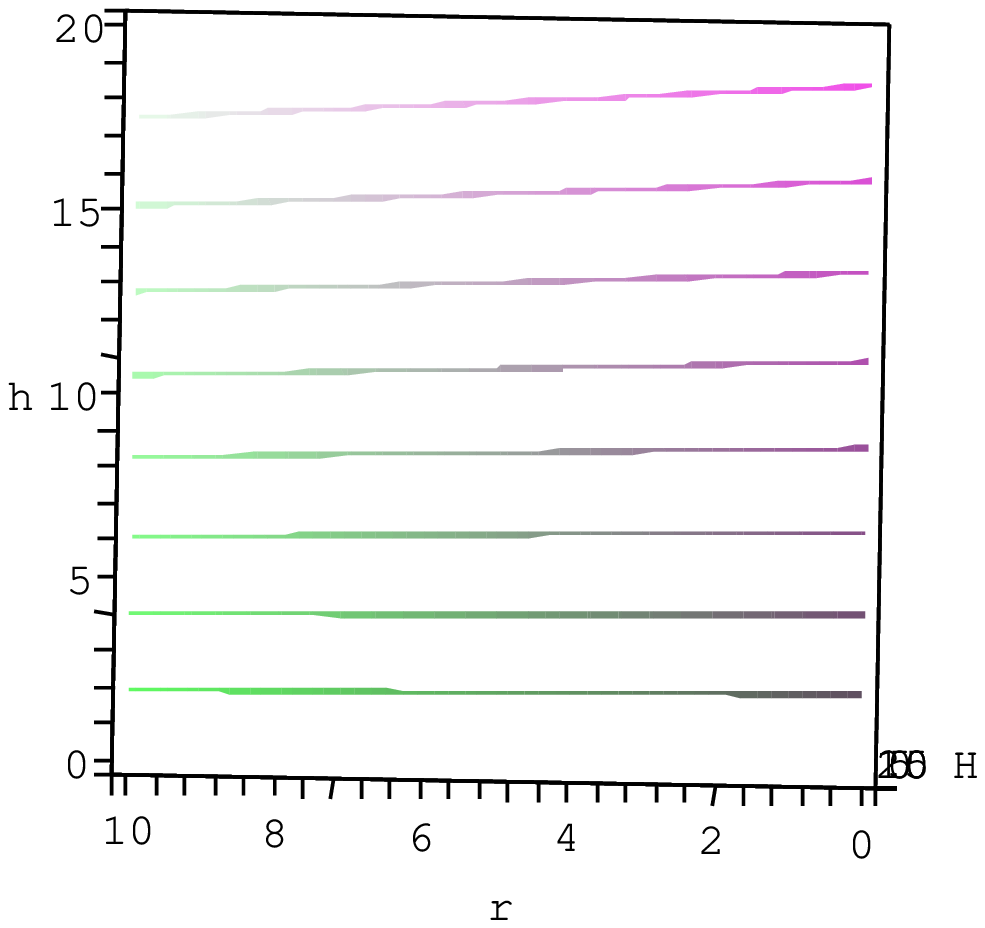}&
\includegraphics[width=.37\textwidth, height=0.37\textwidth]{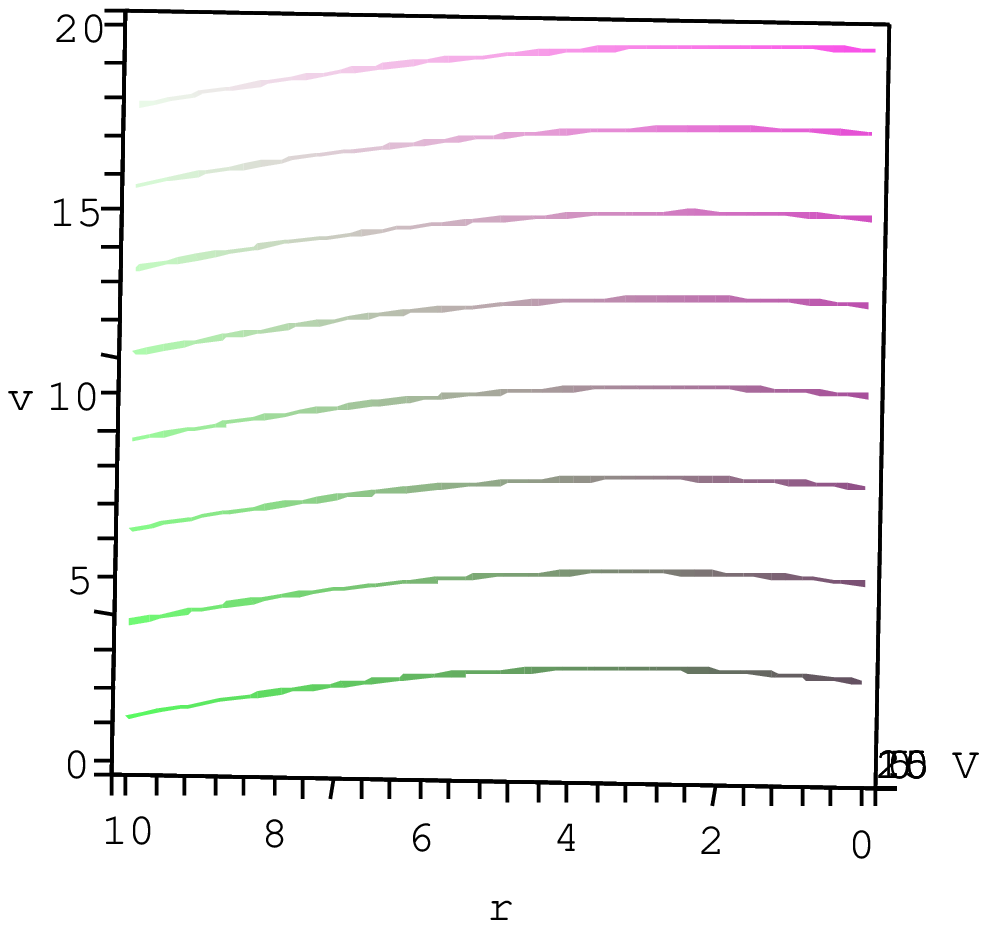}
\end{tabular}
%\medskip

{\small %{\centering\small%\leftskip5em\rightskip5em%\nopagebreak\par
Figure 2.
Horizontal sections, corresponding to dependencies in Figure 1.
They shows, that vertical perceptive geometry differ from Euclidean
geometry in more extent, than horizontal one.
%\hfill
%\par
%}
}

%\medskip

%\refstepcounter{figure}\label{r3d2}
\begin{tabular}{ll}
\includegraphics[width=.37\textwidth, height=0.37\textwidth]{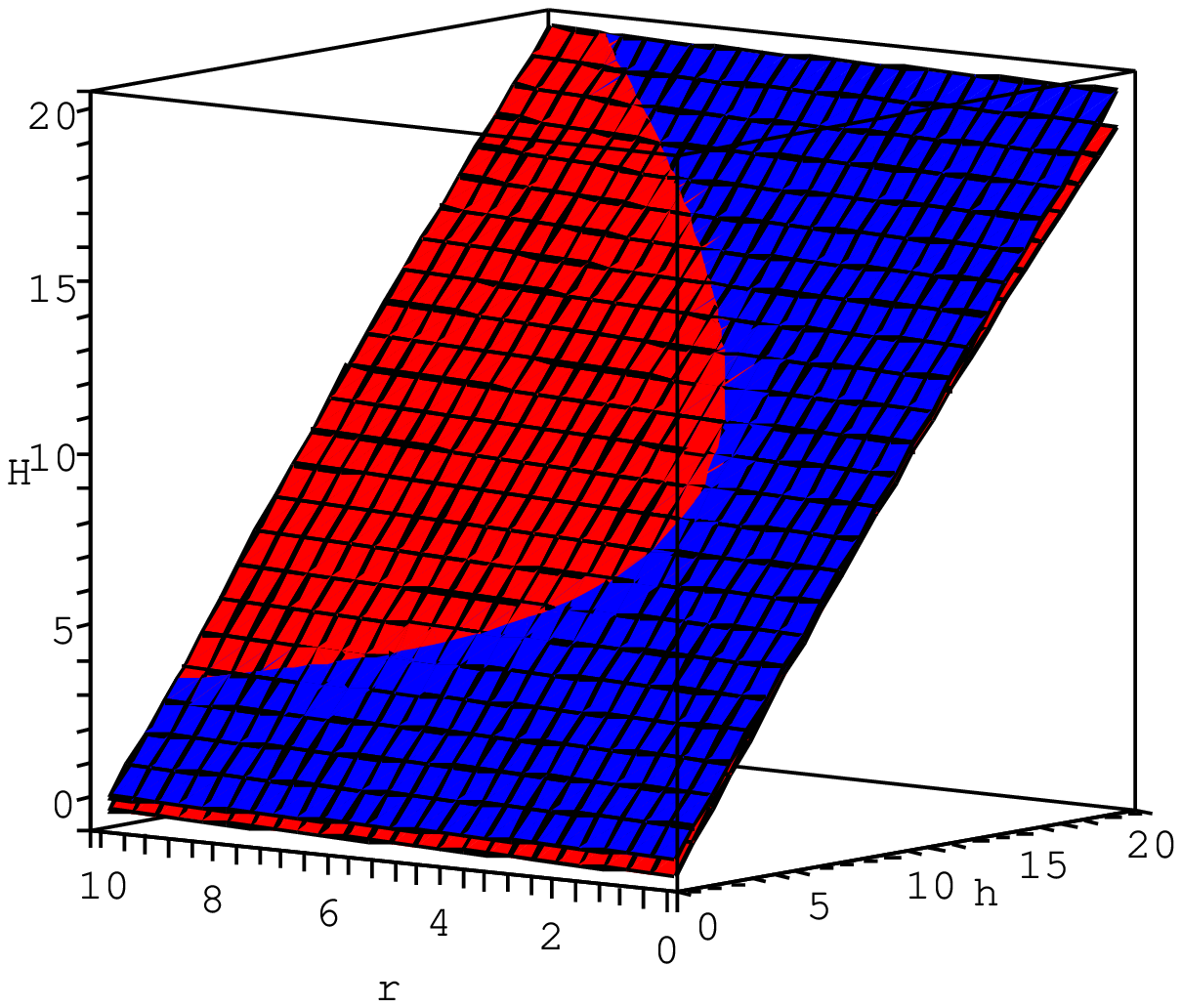}&
\includegraphics[width=.37\textwidth, height=0.37\textwidth]{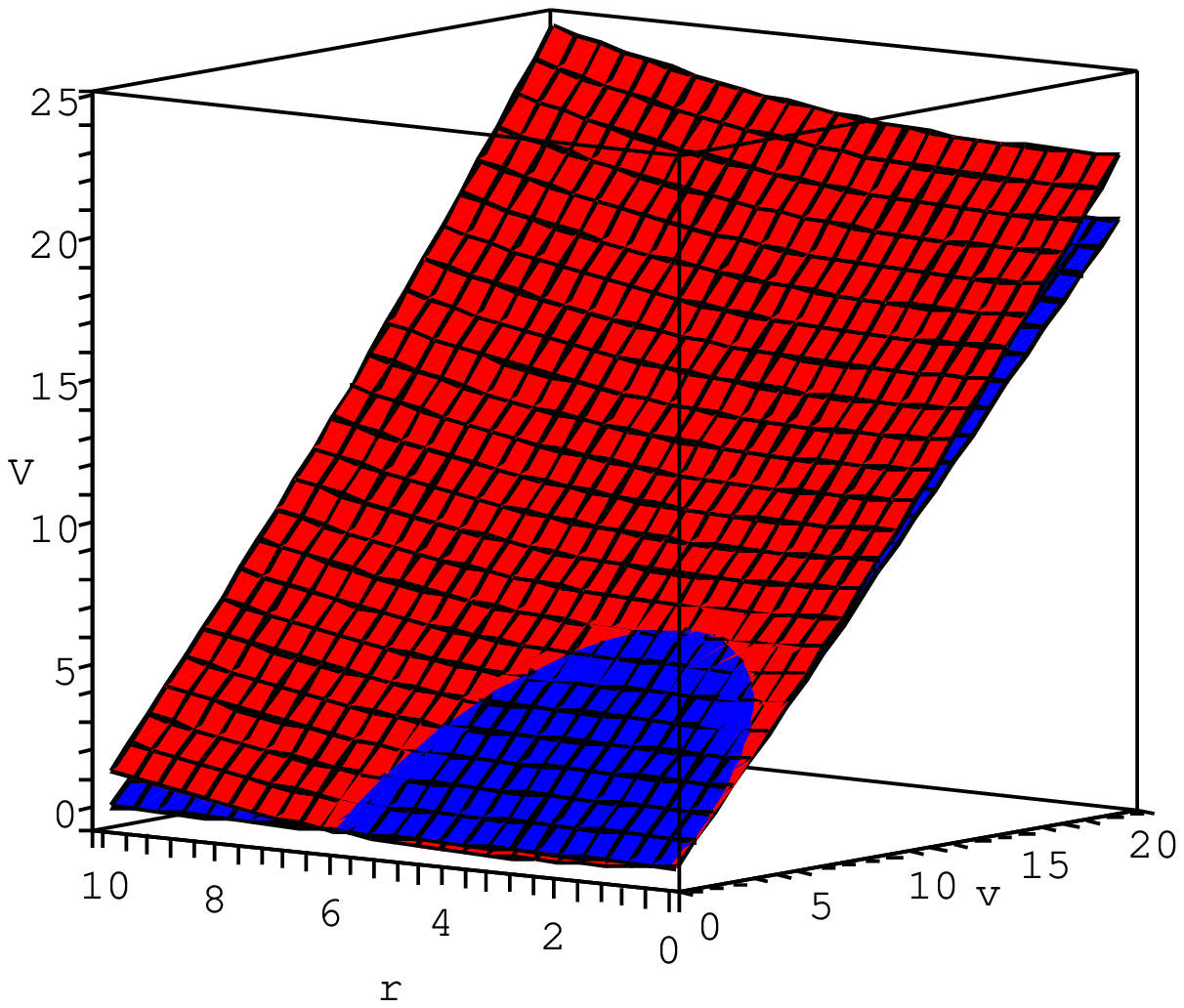}
\end{tabular}
%\medskip

{\small %{\centering\small%\leftskip5em\rightskip5em%\nopagebreak\par
Figure 3.
Sections of experimental dependencies (bright surfaces)
by planes $V=v$ and  $H=h$ respectively (dark surfaces),
describing Euclidean perceptive geometry.
%\hfill
%\medskip%\par
%}
}

\section{Data tables}

{\bf Table 1.} Rectangular's sizes.

\noindent\begin{tabular}{|c|cccccccc|}
\hline
Number & 1 &   2&    3&   4&   5&    6&  7&  8 \\ \hline
Sizes, cm& 5,4$\times$2,2&
0,9$\times$10,2&23,2$\times$2,6&6,5$\times$6,1&12,7$\times$8,1&7,2$\times$19,5&4,5$\times$9,4&1,7$\times$1,7\\
\hline
\end{tabular}

\medskip

{\bf Table 2.} Set of distances in a middle zone.

\noindent\begin{tabular}{|c|cccccccc|}
\hline
Number & 1 &   2&    3&   4&   5&    6&  7&  8 \\ \hline
Distances, m& 0.8&2.2&3.6&4.95&6.25&7.5&8.75&9.9\\
\hline
\end{tabular}
%\newpage
%\section{L┐иїЇ¤ї¤¤Ёа ┐Є│Ї¤Ёа кЁёv°бЁ.}

\medskip

{\bf Table 3.} Averaged experimental data.

\noindent\begin{tabular}{|c|cccccccc|}
\hline
Number  &1&  2&  3&  4&  5&  6&  7&  8 \\ \hline
w-1&   5.2&  5.2&  4.6&  5.4&  5.4&  5.7&  5.6&  5.9 \\
\hline
v-1&   2.0&  2.2&  1.8&  2.3&  2.4&  2.3&  2.5&  2.9 \\
\hline
d-1&   0.7&  1.8&  3.0&  4.0&  5.4&  7.1&  8.8&  9.2 \\ \hline
w-2&   1.0&  0.9&  1.2&  1.2&  1.2&  1.5&  1.4&  1.4 \\ \hline
v-2&   10.1& 8.9&  9.1&  9.6&  10.5& 10.& 10.1& 9.9 \\  \hline
d-2&   0.8&  1.6&  3.1&  4.3&  5.9&  7.2&  6,9&  9.5\\ \hline
w-3&   20.9& 20.4& 21.5& 24.7& 22.6& 24.2& 22.1& 22.2\\ \hline
v-3&   2.6&  2.7&  2.5&  2.6&  2.7&  2.7&  3.4&  2.4\\ \hline
d-3&   0.8&  1.6&  3.0&  5.0&  6.0&  6.2&  8.3&  7.8\\ \hline
w-4&   5.6&  6.3&  6.7&  6.2&  5.6&  6.9&  7.5&  7.7\\ \hline
v-4&   5.6&  6.3&  6.7&  6.2&  5.7&  6.8&  7.5&  7.7\\ \hline
d-4&   0.8&  2.0&  3.3&  4.5&  5.4&  7.1&  7.6&  9.5\\ \hline
w-5&   13.3& 13.8& 14.4& 13.4& 12.8& 16.0& 12.8& 14.2\\ \hline
v-5&   7,7&  8.7&  8.9&  7.9&  7.7&  8.9&  8.1&  9.6\\ \hline
d-5&   0.8&  2.0&  3.3&  4.0&  6.0&  6.6&  8.2&  8.2\\ \hline
w-6&   7.7&  7.9&  5.7&  6.8&  7.0&  6.1&  7.1&  7.5\\ \hline
v-6&   22.6& 23.5& 19.5& 20.3& 21.0& 19.1& 21.0& 23.2\\ \hline
d-6&   0.8&  2.0&  3.0&  4.6&  5.0&  7.1&  7.0&  9.5\\ \hline
w-7&   3.7&  4.1&  4.0&  3.9&  3.8&  4.3&  4.3&  5.7\\ \hline
v-7&   8.5&  8.9&  9.7&  10.5& 8.8&  10.8& 10.8& 13.0\\ \hline
d-7&   0.8&  1.9&  3.2&  4.2&  5.7&  5.8&  8.1&  10.3\\ \hline
w-8&   1.4&  1.6&  1.4&  1.6&  1.8&  1.8&  2.2&  2.0\\ \hline
v-8&   1.4&  1.6&  1.4&  1.5&  1.8&  1.7&  2.2&  2.0\\ \hline
d-8&   0.7&  2.0&  3.2&  4.5&  4.6&  7.0&  9.0&  10.1\\ \hline
\end{tabular}

\bigskip

{\bf Table 4.} Main number values of an affine model of perceptive
geometry.

\noindent\begin{tabular}{|l|cccccccc|}
\hline Number &1&2&3&4&5&6&7&8 \\ \hline
$k_1,\ 10^{-2}$ &9
&5.6  &22  &18.8  &4.8  &-4.1  &13.8  &8.3 \\ \hline $k_2,\ 10^{-2}$
&7.8  &7.7  &3.1  &18.7  &9.1  &-9.4  &38.6  &8.0 \\ \hline $k_3-1$
&-0.022  &-0.084  &-0.146  &-0.083  &-0.137  &-0.1&-0.027  &0.021 \\
\hline $H_0-h,$ cm &-0.5 &0.0  &-2.1  &-1.0  &0.9  &0.0  &-1.0
&-0.5  \\ \hline $V_0-v,$ cm &-0.3  &-0.9  &-0.1  &-0.6  &-0.2  &2.3
&-1.4  &-0.5  \\ \hline $\varepsilon_1,$ $10^{-2}$ &-9  &0  &-9  &-15
&7  &0  &-22  &-29  \\ \hline $\varepsilon_2,$ $10^{-2}$ &-14  &-9
&-4  &-10  &-2  &12  &-15  &-29  \\ \hline $R_0,$ m &-0.48  &-0.25
&0.05  &-0.11  &0.03  &-0.18  &-0.48  &-0.60  \\ \hline $r^\ast_1,$ m
&5.8  &-0.2  &9.8  &5.3  &-18  &-0.7  &7.4  &5.5  \\ \hline
$r^\ast_2,$ m &4.2  &10.5  &3.0  &3.7  &1.9  &24 &3.8  &5.7
\\ \hline $r^\ast_3,$ m &-22  &-3  &0.3  &-1.35  &0.24  &-1.8  &-18
&28  \\ \hline $\sigma$   &0.13  &-0.38  &0.86  &0.005  &-0.9  &-1.3
&-1.8  &0.04  \\ \hline
\end{tabular}

\small
\begin{thebibliography}{99}
\bibitem{geom4}
I.~Ya.~Aref'eva~and~others, "Noncommutative Field Theories and
(Super)String Field Theories", arXiv:hep-th/0111208
\bibitem{geom3}
C.~P.~Bachas, "Lectures on D-branes", arXiv:hep-th/9806199.
\bibitem{simult}
D.Giulini, "Uniquess of Simultaneity", arXiv:gr-qc/0011050
\bibitem{geom2}
M. Green, J. Shwarz, E. Witten,  "Superstring theory", Cambridge University
Press, Cambridge, 1989.
\bibitem{grib}
A.A.Grib, "Bell's inequality violation  and measurements problem in
quantum mechanics", Dubna, UINI, 1992 (In Russian)
\bibitem{hubel}
D.H.Hubel, "Eye, Brain and Vision", Scient. Amer. Libr., New York,
1989.
\bibitem{geom1}
Sh.~Kobayashi,~K.~Nomizu,  "Foundations of differential geometry", v.1,2,
1963.
\bibitem{geom5}
S.S.Kokarev, "Deformational Structures on Smooth Manifolds",
In {\it Trends in Mathematical Physics Research}, Nova Science Publisher,
New York, 2004, 113-154.
\bibitem{kok}
S.S.Kokarev, "Observers, $\mathcal{P}-$structures and  $\mathcal{P}-$geometry",
In Abstracts of 5-th Asian-Pacific conference, PFUR, Moscow, 2001, 37-38.
\bibitem{krylov}
V.Yu.Krylov, "Geometrical representation of datas in psyhological
investigations", Nauka, Moscow, 1990 (In Russian)
\bibitem{kul}
Yu.I.Kulakov and others, "Introduction to physical structures theory
and binary geometrophysics", Archimed, Moscow,  1992 (In Russian)
\bibitem{qrf}
S.N.Mayburov, "Local Reference Frames and Quantum Space Time",
arXiv:hep-th/9708020
\bibitem{mich}
G.G.Michailichenko,
{\it DAN USSR} {\bf 206} 5 (1972) 1056-1058 (In Russian)
\bibitem{mitsk}
N.V.Mitskevich, "Relativistic Physics in Arbitrary Reference Frame",
arXiv:gr-qc/9606051
\bibitem{pav}
M.Pav\v{s}i\v{c}, {\it Nuovo Cimento} {\bf 95 A}  (1986) 297-310
\bibitem{penrose}
R.Penrose, "The Emperor's New Mind", Oxford University Press, 1989.
\bibitem{poincare}
A. Poincare, "Science and Hypothesis", cited from the book "About
Science", Nauka, Moscow, 1990 (In Russian)
\bibitem{spt}
H.Saller, "Space-Time as the Manifold of the Internal Symmetry Orbits
in the External Symmertry", arXiv:hep-th/0103043
\bibitem{cosm}
M.V.Sazhin, "The modern cosmology", URSS, Moscow, 2002 (In Russian)
\bibitem{algebra}
L.A.Skornyakov and others, "General algebra" (2 volumes)\ M.\, Nauka, 1991 (In Russian).
\bibitem{geom6}
A.M.Vinogradov,~I.S.Krasilshchik and others,  "Symmetries and conservation
laws of equations of mathematical physics", Factorial, Moscow, 1997 (In Russian).
\end{thebibliography}
\end{document}